\documentclass[%
 reprint,
superscriptaddress,
nofootinbib,
 amsmath,amssymb,
 aps,
]{revtex4-2}

\usepackage{ulem}
\usepackage{csquotes}
\usepackage{braket}
\usepackage{graphicx}
\usepackage{mathtools}
\usepackage{dcolumn}
\usepackage{bm}
\usepackage{hyperref}
\usepackage{balance}
\hypersetup{colorlinks=true,linkcolor=blue,citecolor=blue,filecolor=blue,urlcolor=blue}
\mathtoolsset{showonlyrefs}
\usepackage{siunitx}
\usepackage{float}
\usepackage{dsfont}

\begin{document}

\title{Data Efficient Prediction of excited-state properties using Quantum Neural Networks}

\author{Manuel Hagelueken}
\email{manuel.hagelueken@ipa.fraunhofer.de}
\affiliation{Fraunhofer Institute for Manufacturing Engineering and Automation IPA, Nobelstraße 12, D-70569 Stuttgart, Germany}

\author{Marco F. Huber}
\affiliation{Fraunhofer Institute for Manufacturing Engineering and Automation IPA, Nobelstraße 12, D-70569 Stuttgart, Germany}
\affiliation{Institute of Industrial Manufacturing and Management IFF, University of Stuttgart, Allmandring 35, Stuttgart, D-70569, Germany}

\author{Marco Roth}%
\email{marco.roth@ipa.fraunhofer.de}
\thanks{Corresponding author}
\affiliation{Fraunhofer Institute for Manufacturing Engineering and Automation IPA, Nobelstraße 12, D-70569 Stuttgart, Germany}

\date{\today}

\begin{abstract}
Understanding the properties of excited states of complex molecules is crucial for many chemical and physical processes. Calculating these properties is often significantly more resource-intensive than calculating their ground state counterparts. We present a quantum machine learning model that predicts excited-state properties from the molecular ground state for different geometric configurations. The model comprises a symmetry-invariant quantum neural network and a conventional neural network and is able to provide accurate predictions with only a few training data points. The proposed procedure is fully NISQ compatible. This is achieved by using a quantum circuit that requires a number of parameters linearly proportional to the number of molecular orbitals, along with a parameterized measurement observable, thereby reducing the number of necessary measurements. We benchmark the algorithm on three different molecules with three different system sizes: $H_2$ with four orbitals, LiH with five orbitals, and $H_4$ with six orbitals. For these molecules, we predict the excited state transition energies and transition dipole moments. We show that, in many cases, the procedure is able to outperform various classical models (support vector machines, Gaussian processes, and neural networks) that rely solely on classical features, by up to two orders of magnitude in the test mean squared error.
\end{abstract}

\maketitle

\begin{figure}[t]
\includegraphics[width=0.48\textwidth,clip, trim= 0cm 0cm 0cm 0cm]{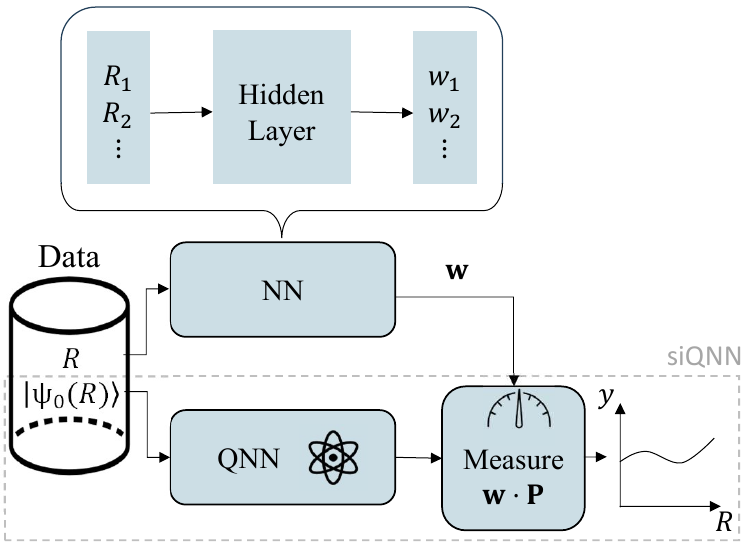}
\caption{\label{fig:method_TotalModel} Schematic overview of the model used in this study. The feature set, i.e., the input to the model is given by $\mathcal{X}=(\ket{\psi_0(R)},R)$, where $\ket{\psi_{0}(R)}$ denotes the molecular ground state and $R$ is the potentially multi-dimensional distance between the atoms. The model is trained to predict various target values $y$ such as the transition energy $\Delta E$ to an excited state. The observable is a weighted sum of Pauli-strings $\mathbf{P}$, where the weights $\mathbf{w}$ are learned by a classical NN [cf. Equation~\eqref{eq:observable_equation}]. Training this model involves a pre-training of the siQNN (dashed box) followed by an end-to-end training of the whole architecture.}
\end{figure}

\section{\label{sec:introduction}Introduction}

Computing molecular properties, such as potential energy surfaces (PES) is a valuable yet often resource-intensive task, making it a highly active research area. A detailed understanding of molecular properties in different geometric configurations is beneficial for various applications in material science, pharmacology, and chemistry~\cite{ExcitedStateOverview2005, PhotostabilityOnDrugs, Voehringer2012}. The ability to predict molecular behavior improves and optimizes processes, leading to more efficient research and applications, such as the discovery of new drugs~\cite{QC_drug_discovery} and the development of catalysts~\cite{QC_catalyst_discovery}.

There are several methods to calculate properties of interest such as ground states and their energies for different molecule sizes with varying precision. For small molecules, full configuration interaction methods~\cite{FCI} can be used to determine the exact ground states and energies in a given basis. Additionally, complete active space self consistent field (CASSCF)~\cite{CASSCF} methods allow the definition of an active space to reduce the problem size and obtain approximate solutions for larger molecules. Despite this and further advances, classical algorithms still struggle with large systems with strongly correlated electrons, as the exact description of these systems scales exponentially with their size~\cite{DrugDesignOnQC}. Therefore, quantum computing algorithms such as the variational quantum eigensolver (VQE)~\cite{VQE} and quantum phase estimation~\cite{QPE} have been studied in the pursuit of a better scaling which is enabled by the exponential size of the quantum Hilbert space. However, these algorithms have their own drawbacks, particularly in the current era of noisy intermediate-scale quantum computing~\cite{NISQ}. Notable problems include a high circuit depth, a large number of measurements~\cite{VQE}, and the necessity for a large overlap of the ansatz with the final state~\cite{QPEHighOverlap}. Consequently, recent studies have focused on combinations of classical and quantum algorithms, aiming to leverage the strengths of both approaches~\cite{ceroni2023generatingapproximategroundstates,robledomoreno2024chemistryexactsolutionsquantumcentric}.

Most methods, especially for quantum computing, are tailored to obtain ground states and their energies. However, for many chemical and physical processes, such as photochemical reactions, energy transfer, and the absorption and emission of light, excited states are of significant interest. An accurate calculation of these states and their energies is much more challenging because they are usually composed of more electronic configurations than the ground state~\cite{excited_states_configurations}. Additionally, electron correlations are more pronounced, and they can have different spin multiplicities and symmetries. Furthermore, it is challenging to correctly differentiate between individual states due to effects such as (avoided) crossings~\cite{ExcitedStateChallenges}. There are several extension to the commonly used quantum computing algorithms to calculate excited states, such as variational quantum deflation~\cite{VQD} or subspace-search VQE~\cite{ssVQE}, but since these often are based on VQE they usually suffer from the same drawbacks.

To address these challenges we propose a quantum machine learning (QML)\cite{Cerezo2022} model that directly operates on the quantum mechanical ground state on a quantum computer (QC). The quantum state is measured using only commuting operators drastically reducing the overhead on the QC. The model includes a quantum neural network (QNN) and a classical neural network (NN) designed to efficiently calculate excited-state properties\footnote{We use \enquote{excited-state property} not only for intrinsic properties of the excited state but also to excited state related properties such as transition energies.} for different geometric configurations of a given molecule. The architecture of the QNN is constructed to explicitly account for the symmetries of the ground state~\cite{PRXQuantum.4.010328,Schatzki2024,le2023symmetryinvariantquantummachinelearning}. The proposed model is able to predict these properties with only a few training data points which avoids costly excited state simulations. The overall procedure is sketched in Figure~\ref{fig:method_TotalModel}.

To assess the performance of our method, we compare it to well-known classical methods. We benchmark these models on three different molecules for various excited states. We calculate the transition energy of the ground state to excited states ($\Delta E$) and the $L_{2}$-norm of the transition dipole moments (TDM) between the ground state and excited states for varying interatomic distances. We find that using the ground state as an additional feature significantly improves the prediction quality, especially in the low-data regime.

In a related work~\cite{ClassicalShadowExcitedEnergies} the first transition energies and TDMs are predicted by performing measurements with different observables on the ground state after an non-parametrized entangling layer. The measurement values are used as input for an NN that outputs the target functions. In~\cite{DNNForExcitedStates} a deep neural network determines the parameters of a QNN that prepares excited states using variational quantum deflation or the subspace search VQE. While these approaches share a similar rationale regarding the benefits of interpolation in this context, their models are trained with a significantly larger amount of training data (about a factor of 5 times the training data points our models uses) reaching similar accuracy as our model. Therefore, they do not compare themselves to classical interpolation methods, as we do.

The remainder of this work is organized as follows. In Section~\ref{sec:methods}, we motivate and discuss the construction of the model and its training procedure. In Section~\ref{sec:results}, we demonstrate the efficacy of the model in predicting excited-state properties from the ground states of LiH, $\mathrm{H}_2$, and rectangular $\mathrm{H}_4$, and provide evidence for its data efficiency compared to several other models. Finally, in Section~\ref{sec:discussion}, we discuss our findings, the limitations of our model and study design, and propose directions for further research.

\section{methods}
\label{sec:methods}
In this section, we define the problem and present the QML model. We describe how the ground state of a molecule can be obtained on a QC utilizing the Jordan-Wigner transformation. From this, we discuss relevant symmetry considerations and use this insight to tailor our model to the task of predicting excited-state properties from the ground state.

\subsection{\label{sec:methods_WF}Molecular States on a QC}
The Fermionic Hamiltonian of a molecule in second quantization is of the form
\begin{equation}
H = \sum_{pq} h_{pq} a_p^\dagger a_q + \frac{1}{2} \sum_{pqrs} h_{pqrs} a_p^\dagger a_q^\dagger a_r a_s\,,
\label{fermionic_hamiltonian}
\end{equation}
where $h_{pq}$ and $h_{pqrs}$ are the Coulomb overlap and exchange integrals, and $a_i^\dagger$ and $a_i$ are Fermionic creation and annihilation operators. To make this Hamiltonian computationally tractable, we restrict the orbitals to an active space containing only the most important contributions to our quantity of interest. To analyze molecules on a QC, the Fermionic operators need to be mapped to qubit operators. In this work, we use the Jordan-Wigner mapping due to its straightforward physical interpretation~\cite{JW_transformation}. Using this mapping, the creation operators can be expressed as 
\begin{equation}
a_p^\dagger = \frac{1}{2} \left( \mathrm{X}_p - i\mathrm{Y}_p \right) \prod_{j=1}^{p-1} \mathrm{Z_j}\,,
\end{equation}
where $X$, $Y$, and $Z$ are the Pauli operators of the $i$-th qubit. The annihilation operator is given analogously. From this we obtain a qubit Hamiltonian
\begin{equation}
H_{\text{qubit}} = \sum_{i=0}^K c_i P_i\,,
\label{eq:qubit_hamiltonian}
\end{equation}
where the coefficients $c_i$ are determined by $h_{pq}$ and $h_{pqrs}$ and $P_i$ are Pauli-strings determined by the mapping.

With this mapping, each spin-orbital is represented by a qubit, where an excited qubit state corresponds to an occupied orbital. Since electrons can be either spin-up ($\alpha$-electron) or spin-down ($\beta$-electron), $2n_{\mathrm{orb}}$ qubits are needed to represent a molecule with $n_{\mathrm{orb}}$ orbitals. The exact ordering of qubits to spin-orbitals can vary. In this work the first $n_{\mathrm{orb}}$ qubits represent one type (e.g., $\alpha$) of electrons, and the second $n_{\mathrm{orb}}$ qubits represent the other type of electrons. Furthermore, the first qubit of both spin-subsets, i.e., qubit 0 and qubit $n_{\mathrm{orb}}$, correspond to the first orbital (lowest energy configuration), the second to the second orbital, and so on. This ordering of orbitals by energy is established near the equilibrium configuration and is preserved throughout the dissociation curve to enhance the generalization performance of the QNN across the dissociation.

\begin{figure}[tb]
\includegraphics[width=0.48\textwidth,clip, trim= 0cm 0cm 0cm 0cm]{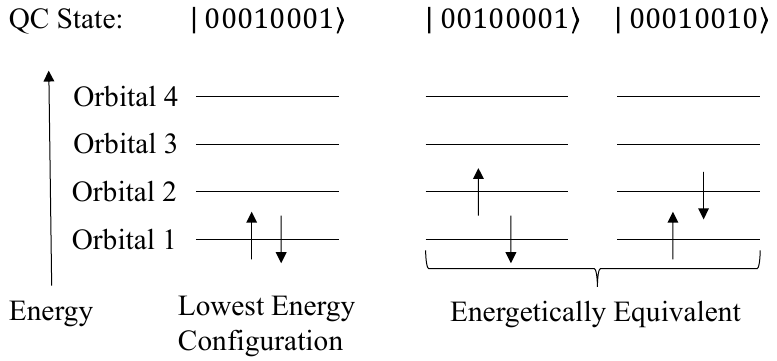}
\caption{\label{fig:method_AnalyseWF} A simple illustration of the symmetry of several electronic ground state configurations for $\mathrm{H}_{2}$ with four considered orbitals prepared on the QC using the Jordan-Wigner mapping.}
\end{figure}

We consider molecules in an isolated environment, resulting in a spin-degeneracy between the $\alpha$ and $\beta$ orbitals. Figure~\ref{fig:method_AnalyseWF} illustrates three spin configurations and their corresponding Jordan-Wigner mapped states for the example of $\mathrm{H}_2$. Near the equilibrium the lowest energy configuration has the highest amplitude in the ground state. However, higher energy configurations also contribute to the ground state but with smaller amplitudes. Each configuration is associated with a computational basis state on the QC. The squares of the amplitudes of these computational basis states represent the probabilities of electrons occupying the corresponding spin-orbitals. As the molecule dissociates, the shapes of the orbitals change, leading to variations in the occupation probabilities of the spin orbitals. Consequently, the amplitudes of the computational basis states in the ground state change accordingly. In the following section, we design a QNN to read and process these amplitudes and their variations to derive the target value $y$ (in our case an excited-state property in a given geometric configuration) using training data points. For additional theoretical justification for why it is possible to predict properties of excited states from the ground state, we refer to~\cite{ClassicalShadowExcitedEnergies}.

\begin{figure*}[htbp]
\includegraphics[width=0.9\textwidth,clip, trim= 0cm 0cm 1.6cm 0cm]{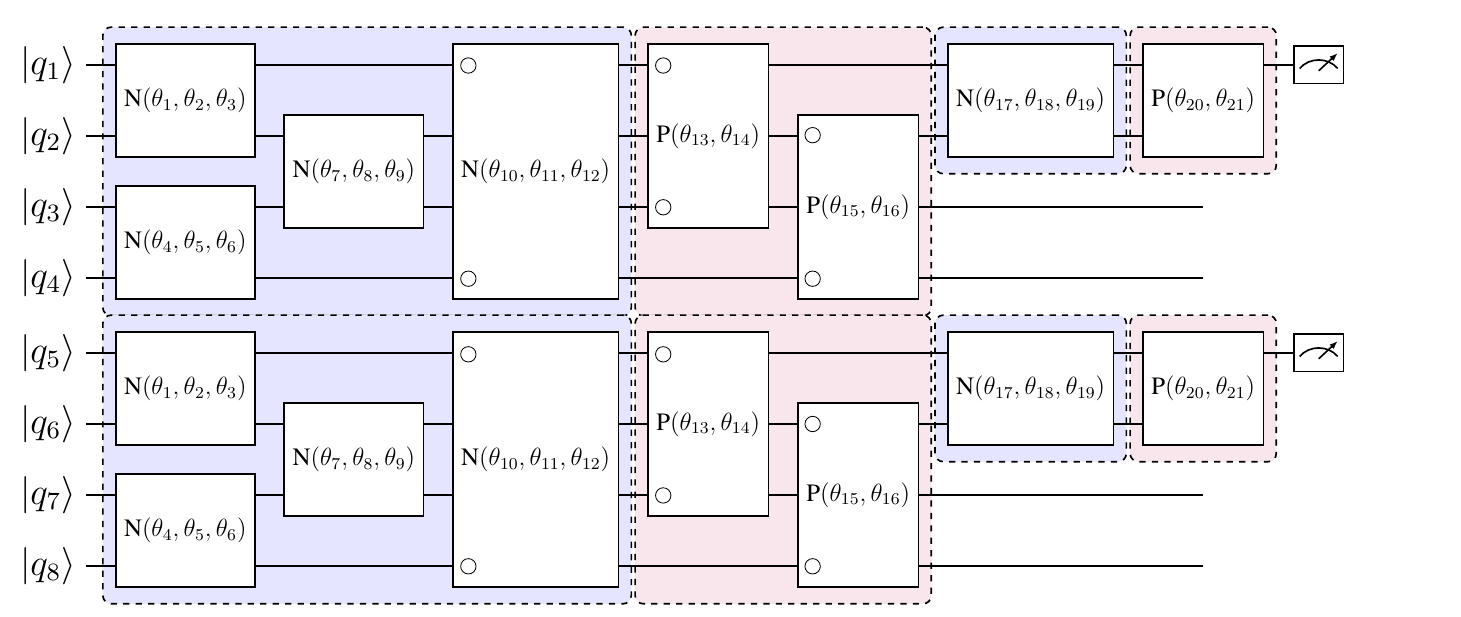}
\caption{\label{fig:method_QnnAnsatz} An example of the structure of the QNN ansatz for $\mathrm{H}_2$ with 4 orbitals and 2 electrons. Here, $\theta_{1}$--$\theta_{21}$ are trainable parameters and $q_1$--$q_8$ denote the qubits. The two-qubit gate $\mathrm{N}(\theta_i,\theta_j,\theta_k)=\exp[i(\theta_iXX+\theta_jYY+\theta_kZZ)]$ can realize every two qubit unitary~\cite{optimal_two_qubit_gate} and $P(\theta_n, \theta_m)$ is a two qubit entangling gate that acts as a pooling operation~\cite{QCNN}. When gates act on non-neighboring qubits, circles in the gates mark the qubits on which the gates are applied.}
\end{figure*}

\subsection{\label{sec:methods_model_ansatz}Model Ansatz}
The intuitive approach to derive the transition energy $\Delta E$ to an excited state from the ground state would be to find the unitary excitation operator $U_j^{\rm exc}$ that transitions the ground state to the $j$-th excited state and measuring the Hamiltonian $H$ of the system in the excited state. From the result, one can subtract the measurement of $H$ in the ground state to obtain $\Delta E$. However, there are three problems with this approach that we address with the specific design choices of our model.

Firstly, the structure of $U_j^{\rm exc}$ is generally not known, and an exact derivation would introduce a significant computational overhead. Therefore, we aim to model its action with a machine learning approach consisting of a QNN with $U_{\rm QNN}$ acting on the ground state.

Secondly, in general, a large number of shots are required to derive $\Delta E = E_j - E_{0}$ by measuring $E_0$ and $E_j$ before subtracting. This is because for larger molecules $\Delta E$ can be on a lower energy scale than $E_0$ and $E_j$ and if one wants to know $\Delta E$ to a certain accuracy, $E_0$ and $E_j$ need to be derived to a larger accuracy, which leads to a larger number of shots required, compared to a direct measurement of $\Delta E$. Therefore, we aim to directly approximate the transition energy Hamiltonian $\tilde{H}$, i.e.,
\begin{alignat}{2}
\Delta E &= \langle \psi_0 | \left(U_j^{\rm exc}\right)^\dagger H U_j^{\rm exc} | \psi_0 \rangle - \langle \psi_0 | H | \psi_0 \rangle \notag \\
&= \langle \psi_0 |\left(U_j^{\rm exc}\right)^\dagger H U_j^{\rm exc}  -  H | \psi_0 \rangle \equiv\langle \psi_0 | \tilde{H} | \psi_0 \rangle\,,
\label{eq:delta_E_htilde}
\end{alignat}
where $U_{\mathrm{diag}}$ is the unitary that diagonalizes $\tilde{H}$, so that we can measure its eigenvalues in the computational basis with the measurement $D$.
Since the Hamiltonian can be expressed as a linear combination of many non-commuting Pauli-strings [see Equation~\eqref{eq:qubit_hamiltonian}] the same holds true for $\tilde{H}$ in Equation~\eqref{eq:delta_E_htilde}, because $U_j^{\rm exc\dagger} H U_j^{\rm exc}$ is also a linear combination of (potentially different) Pauli-strings. Theoretically, it is possible to derive this unitary operation $U_{\mathrm{diag}}$ that diagonalizes $\tilde{H}$, so that we can measure its eigenvalues (the $\Delta E$) in the computational basis with
\begin{equation}
O(\mathbf{w}) = w_1 \mathds{1} + w_2 (Z_0 + Z_{n_\mathrm{orb}+1}) + w_3 Z_1 Z_{n_\mathrm{orb}+1} + \ldots\,,
\label{eq:observable_equation}
\end{equation}
where $\mathds{1}$ is the identity. Note that this observable explicitly respects the spin symmetry discussed in Section~\ref{sec:methods_WF}. Since we do not know the exact structure of $\tilde{H}$ we leave it to the QNN to rotate into its eigenbasis and to derive the weights $w_i$ of the resulting measurements in this basis. We thus approximate $\Delta E$ by measuring
\begin{equation}
\Delta E \approx \bra{\psi_0}U_{\rm QNN}^\dagger O(\mathbf{w}) U_{\rm QNN}\ket{\psi_0}\,.\label{eq:delta_e_approx}
\end{equation}
Depending on the complexity of the problem, we might not find the exact eigenbasis of the full $\tilde{H}$ but only its leading contributions. 

The third challenge is that the coefficients in $H$ and therefore also in $\tilde{H}$ generally depend on the geometric configuration $R$ of the given molecule. In this work, $R$ is a scalar describing the distance between the atoms of the molecule, but in general $R$ can be a multidimensional vector. Although one might hope that the $R$ dependency could be negligible or that the QNN could project into an $R$-independent space, we explicitly consider this $R$ dependence by introducing a small classical NN. This NN takes $R$ as input and outputs the weights $w_i$ in Equation~\eqref{eq:observable_equation}. The linear combination of these operators constitutes the output of the model. Therefore, the NN is designed to learn and account for the dependence of $R$ in $\tilde{H}$. An example of the overall structure of the model is depicted in Figure~\ref{fig:method_TotalModel}. All the same arguments that lead to Equation~\eqref{eq:delta_e_approx} can be used to derive general matrix elements connecting the ground state to the $j$-th excited state of an operator of interest $V$, i.e.,
\begin{align}
\| V_{i,j}\|_2 &= \| \langle \psi_j | V | \psi_{i} \rangle \|_{2} =\sqrt{\langle \psi_i | V^\dagger | \psi_{j} \rangle \langle \psi_j | V | \psi_{i} \rangle}
\\
& \equiv \langle \psi_i | \tilde{V} | \psi_i \rangle\,,
\label{eq:matrix_element_derivation}
\end{align}
where we have introduced $\tilde{V}= \sqrt{V^\dagger | \psi_{j} \rangle \langle \psi_j | V}$ in the last equation. In this work, we are primarily interested in obtaining the $L_2$-norm of the matrix element which is denoted by $\|\cdot|\|_2$ in the equation above. Note that other expectation values and matrix elements can be approximated similarly. For the structure of the QNN we aim for expressiveness, so that the QNN can implement the necessary operations, and efficient trainibility with a useful bias towards the overall operations it is meant to perform, so that the correct operations are chosen. To integrate the molecular symmetry into the QNN, we consider only spin-symmetric operations. This is achieved by using the same operations on the first $n_\mathrm{orb}$ and last $n_\mathrm{orb}$ qubits. 

To enhance trainability, the structure of the QNN is partially inspired by quantum convolutional neural networks, which are known to avoid barren plateaus and have a circuit depth that scales logarithmically~\cite{QCNN}. Together with the given symmetry restriction, the number of parameters is further reduced, scaling linearly with the number of orbitals considered. Specifically, when pooling to two qubits, the number of parameters is $n_{\mathrm{params}} = 8(n_{\mathrm{orb}}-1)-3$ (see Appendix~\ref{sec:appendix_QNN_parameter_scaling}). The QNN consists of alternating two-qubit operation layers. The first layer (N) is designed for maximal expressivity between neighboring qubits and is intended to implement rotations between adjacent orbitals inspired by ADAPT-VQE structural approaches (compare, e.g.~\cite{fixed_adapt_vqe_disso_ansatz}), while the second layer (P) is a pooling layer designed to entangle the given qubits. After each pooling layer, half the qubits are discarded and the information is pooled onto the remaining qubits. These operational layers are repeated until only a small number of qubits remain. Note that for large systems, this process could be repeated until the number of remaining qubits equals the number of electrons in the system. In this case, additional contributions in Equation~\ref{eq:observable_equation} are further Pauli-strings consisting of $Z$-operators, respecting the given symmetry, and acting on qubits which are left in the circuit after the pooling operations (indicated by the dots in the equation). Figure~\ref{fig:method_QnnAnsatz} shows an example of such a QNN for $\mathrm{H}_2$ with four orbitals. We call the resulting symmetry-invariant QNN combined with an NN \enquote{siQNN-NN}. Quantum convolutional neural networks are known to be simulatable classically~\cite{QCNN_simulatable}. The relevance of these findings, namely the fact that this is not trivially possible but might be an interesting further research direction, is discussed in Section~\ref{sec:discussion}. Note that we have chosen the particular QNN architecture in Figure~\ref{fig:method_QnnAnsatz} for its computational efficiency. However, the proposed method of letting a QNN operate directly on ground states is not restricted to this specific choice. 

\begin{figure*}[t]
\includegraphics[width=1.0\textwidth,clip, trim= 0cm 0cm 0cm 0cm]{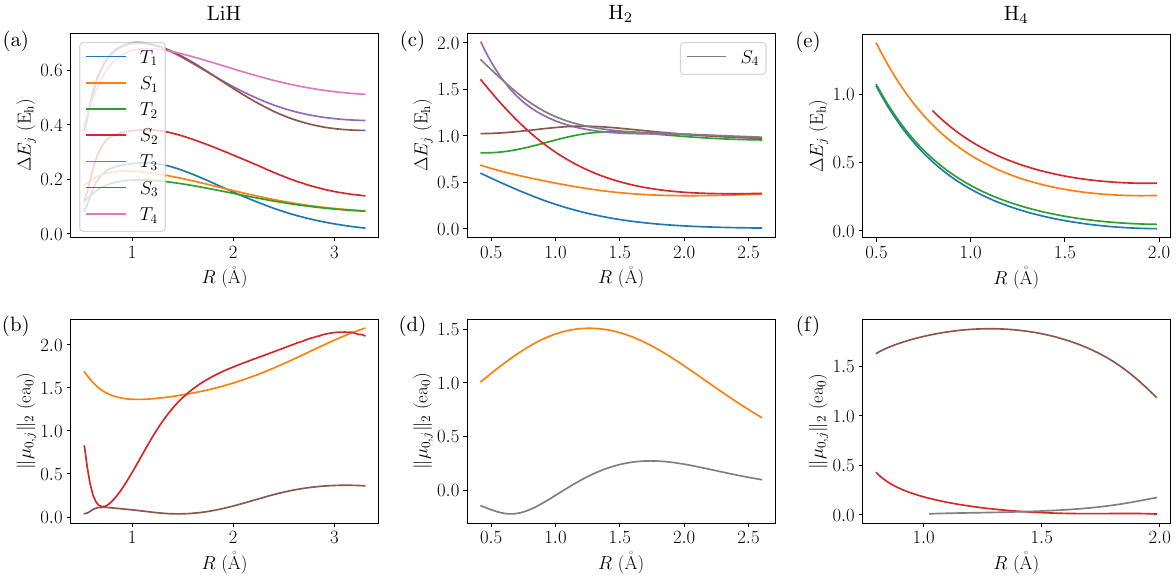}
\caption{\label{fig:method_target_functions} The target functions for LiH in (a) and (b), for $\mathrm{H}_2$ in (c) and (d), and for $\mathrm{H}_4$ in (e) and (f) in the 6-31G basis. In (a), (c), and (e) the transition energies $\Delta E_j$ for the four energetically lowest excited singlet and triplet states, that do not exhibit avoided crossings, are shown in Hartree $\rm E_{\rm h}$. In (b), (d), and (f) the non-zero TDMs of those states are shown in units of elementary charge e and the Bohr radius $\rm{a}_0$. All values are obtained via exact diagonalization of Equation~\eqref{eq:qubit_hamiltonian}.}
\end{figure*}

\subsection{\label{sec:methods_training}Training}
We train the siQNN-NN model in two steps. In the first step, only the symmetry-invariant QNN (siQNN) is trained (gray dashed outline in Figure~\ref{fig:method_TotalModel}). In this step, the weights $w_i$ [cf. Equation~\eqref{eq:observable_equation}] are not parameterized by an NN but rather learned directly through  gradient descent. The aim of this pre-training is to navigate the parameters of the siQNN towards a favorable region in the reduced optimization landscape, i.e., the landscape of the siQNN without the NN. This is followed by an end-to-end training of the siQNN-NN model. Here, the QNN is initialized with the pre-trained parameters while the NN parameters are initialized randomly.

In the best case, the siQNN parameters are already optimal after pre-training, and the model output is similar to the target function, because a good approximation to the eigenbasis of $\tilde{H}$ is found. Then, the NN only fine-tunes the siQNN output towards the target function in regions where it is necessary to account for minor effects of the $R$ dependence and potentially small contributions from a small term in $\tilde{H}$. However, in some cases (for example, the target function $\|\boldsymbol{\mu_}{0,S_2}\|_2$ for LiH, see Figure~\ref{fig:method_target_functions}) the siQNN is unable to adequately describe the target function on its own, due to its restricted expressiveness. Then, the pre-training might converge to parameters that are sub-optimal for the complete siQNN-NN model. In this case, the combined training becomes more important and the NN is used to account for the pronounced $R$ dependence and additional term contributions in $\tilde{H}$.

In both training steps, we use an Adam~\cite{adam} optimizer with mean squared error (MSE) as a loss function. Training is terminated if one of three criteria is met: 1) if the training loss decreases below a predefined target loss $\mathcal{L}_{\rm t}$, 2) if the training loss does not improve over a certain number of iterations, or 3) if a maximum number of iterations is reached.

\section{results}
\label{sec:results}
In this section, we evaluate the performance of the model on three exemplary molecules (LiH, $\mathrm{H}_2$, and $\mathrm{H}_4$). We perform regression on a variety of transition energies and dipole moments and compare the result to other classical ML models.

\subsection{\label{sec:results_datasets}Datasets}
In selecting appropriate example molecules to benchmark the model, we try to satisfy a trade-off between sufficient complexity and reasonable computational costs. We therefore choose LiH which, with $n_{\mathrm{orb}}=5$, is small enough to be simulated, trained, and extensively studied while already exhibiting interesting electronic characteristics. Furthermore, to analyze the scalability of our model, we also benchmark it on $\mathrm{H}_2$ with $n_{\mathrm{orb}}=4$ and rectangular $\mathrm{H}_4$ with $n_{\mathrm{orb}}=6$. 

For each molecule, we calculate the corresponding Hamiltonian Equation~\eqref{fermionic_hamiltonian} in an active space of $n_{\rm orb}$ orbitals using CASSCF with the \enquote{6-31G} basis set using PySCF~\cite{pyscf}. We then obtain the respective qubit Hamiltonian Equation~\eqref{eq:qubit_hamiltonian} using Qiskit~\cite{qiskit}. The qubit Hamiltonian is diagonalized to derive the eigenstates $\ket{\psi_j}$ and their energies $E_j$. From these we construct a regression dataset $\mathcal{D}=(\mathcal{X},\mathcal{Y})$ for each molecule with $y\in\mathcal{Y}\subset\mathbb{R}$. The features consist of the ground states $\ket{\psi_0(R_k)}$ and the corresponding atomic distance $R_k$, i.e., $\mathcal{X}=(\mathcal{S}, \mathcal{R})$, where $\mathcal{S}=\lbrace\ket{\psi_{0}(R_0)},\dots, \ket{\psi_{0}(R_M)}\rbrace\subset \mathbb{C}^{2^{2n_{\rm orb}}}$ and $\mathcal{R}=\lbrace R_0,\dots,R_M\rbrace\subset\mathbb{R}$ with the number of data points $M$. In our experiments, we fix $M=100$ if not mentioned otherwise and select data points with equal spacing in $R$. 

In our benchmarks, we consider two different excited-state properties for the targets $y$: transition energies and TDMs. The transition energies are given by $\Delta E_j = E_j-E_{S_0}$, where $j\in\lbrace S_1, S_0,\dots,T_1,T_2,\dots\rbrace$ labels the corresponding excited singlet ($S_i$) and triplet states ($T_i$). For the TDMs, we calculate the $L_2$ norm of the matrix element connecting the ground state to the $j$-th excited state. This corresponds to choosing $V=\boldsymbol{\mu}$ in Equation~\eqref{eq:matrix_element_derivation}, where $\boldsymbol{\mu}$ is the dipole moment operator. In calculating the TDMs, we take into account possible sign changes to ensure differentiability across the chosen $R$ spectrum.

For each molecule, we calculate the first four excited singlet and triplet state energies and the corresponding non-zero TDMs, see Figure~\ref{fig:method_target_functions}. The upper row displays the transition energies, while the lower row shows the TDMs. The functions exhibit different degrees of complexity. Thus, it can be expected that different numbers of data points are required to describe each function with similar accuracy. The TDMs for LiH are comparatively complex, whereas the transition energies for $\mathrm{H}_4$ and partially for $\mathrm{H}_2$ are relatively simple. Furthermore, several target functions for a given molecule differ mainly by a constant offset, and therefore the performance of the models on these functions is similar. To avoid learning problems, we only retain the states in which the entire $R$ region, from repulsion to dissociation, does not exhibit obvious avoided crossings~\cite{ceroni2023generatingapproximategroundstates}. During training, we scale $R$ and all target values to the interval $[-1,1]$.

In the following experiments, we assume that the molecular states on the QC are provided by some quantum algorithm. In the NISQ era, this could, for example, be done using VQE. To isolate the performance of the model from potential impurities in the input quantum states, we initialize the quantum register directly to state vectors obtained by diagonalizing the qubit Hamiltonian Equation~\eqref{eq:qubit_hamiltonian}.

We compare the performance of the siQNN-NN model to well-known classical models with a reduced feature space, i.e., $\overline{\mathcal{X}}=\mathcal{R}$. The models are trained only on this reduced feature space because the size of the state vector of a molecular ground state is $2^{2n_{\mathrm{orb}}}$. Even if the amplitudes were available classically (e.g., by state tomography), including the molecular states in $\mathcal{X}$ would make the feature dimension more than an order of magnitude larger than the size of the training dataset used in this study, preventing generalization for these models. However, in Section~\ref{sec:discussion} we argue that it might be possible to construct a purely classical model that manages to extract valuable information from the large state vector of the molecular states, but this would require a much more sophisticated ansatz and would therefore be an interesting follow-up research topic.

\begin{figure*}[t]
\includegraphics[width=0.98\textwidth,clip, trim= 0cm 0cm 0cm 0cm]{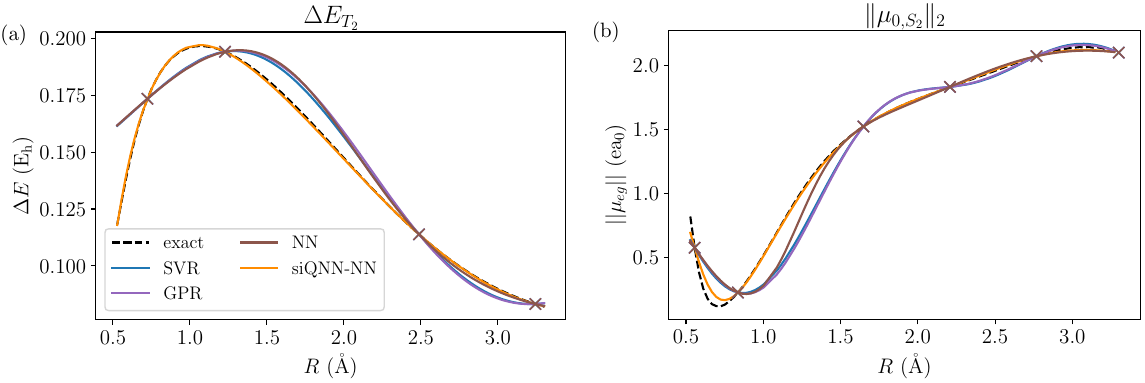}
\caption{\label{fig:results_BO-6-31G-LiH_best_example} The predictions (solid lines) of (a) $\Delta E_{T_{2}}$ and (b) $\|\boldsymbol{\mu_}{0,S_2}\|_2$ for LiH for various models trained on specific trainings datasets. In (a) four training data points (marked by crosses) and in (b) six training data points are used. The exact target functions are represented by dashed lines. In (a) the best MSE score on the test dataset is achieved by the siQNN-NN, with a value of $\qty{2.6e-7}{\hartree}^2$. The second best test MSE score is obtained by the SVR, with a value of $\qty{9.1e-5}{\hartree}^2$. In (b) the best MSE score on the test dataset is achieved by the siQNN-NN, with a value of $\qty{8.3e-4}{\hartree}^2$. The second best test MSE score is obtained by the NN, with a value of $\qty{3.7e-3}{\hartree}^2$.}
\end{figure*}

\subsection{\label{sec:results_model_details}Model Details}
For the siQNN-NN, we use an NN with a single hidden layer comprising 41 parameters with a \enquote{tanh} activation function for the hidden layer. The NN is implemented using PyTorch~\cite{pytorch2024}, while the QNN is implemented in sQUlearn~\cite{squlearn2023} and Qiskit, and then integrated into the model and trained using PennyLane~\cite{pennylane} with state vector simulations. In Appendix B, we additionally show calculations for simulations including shot noise and find promising results. Pre-training is performed with a large learning rate of 0.5 and a maximum of 1,000 iterations to broadly explore the optimization landscape and prevent getting stuck in local minima. The subsequent training of the siQNN-NN is conducted with a learning rate between 0.005 and 0.05 and a maximum of 2,000 iterations to enable smooth convergence. The performance of the model learning on the ground state derived after pre-training labeled siQNN is also compared to the performance of the classical models in the following.

We compare the performance of these models to a purely classical NN that has the same structure as the siQNN-NN, but without the quantum component. The model is trained on the reduced feature space $\overline{\mathcal{X}}$. To ensure comparability, the number of parameters in the NN is equal to the total number of parameters in the siQNN-NN (the sum of parameters in the QNN and the NN). Choosing the same architecture as in the siQNN-NN allows us to isolate the contributions of the classical NN and QNN to make sure that the performance of the siQNN-NN does not solely rely on the classical part of the model. This model is trained with an ADAM optimizer until convergence (if over 1,000 iterations with 5 different learning rates between 0.01 and 0.1 the target loss is not improving) or until the target training loss $\mathcal{L}_{\rm t}$ is reached. Additionally, a support vector regressor (SVR) and a Gaussian process regressor (GPR) with a radial basis function kernel are implemented using scikit-learn~\cite{scikit_learn}. Their hyperparameters are optimized using a grid search, where the performance of the hyperparameters is evaluated on the training dataset.

Because we operate on data that is very computationally expensive, especially for large molecules, we are interested in a low-data regime. In this regime, a separate validation dataset can not be meaningfully defined because having a validation set with only a couple of data points would cause significant information leakage from the validation set. Furthermore, a separate validation dataset is not required, as there is no need for hyperparameter optimization due to the low influence of the hyperparameters on the accuracy. However, without a validation dataset, the training of the (Q)NN-based model cannot be stopped when the performance on the validation dataset worsens and overfitting occurs, but instead it is stopped when $\mathcal{L}_{\rm t} = \qty{1e-6}{\hartree}^2$ is reached on the training dataset. We choose this particular value as it is just below chemical accuracy.  Recent work has found that QML methods generally avoid systematic overfitting of the training data, making the low data regime potentially interesting for QML methods in general~\cite{bowles2024betterclassicalsubtleart}.

\begin{figure*}[t]
\includegraphics[width=1.0\textwidth,clip, trim= 0cm 0cm 0cm 0cm]{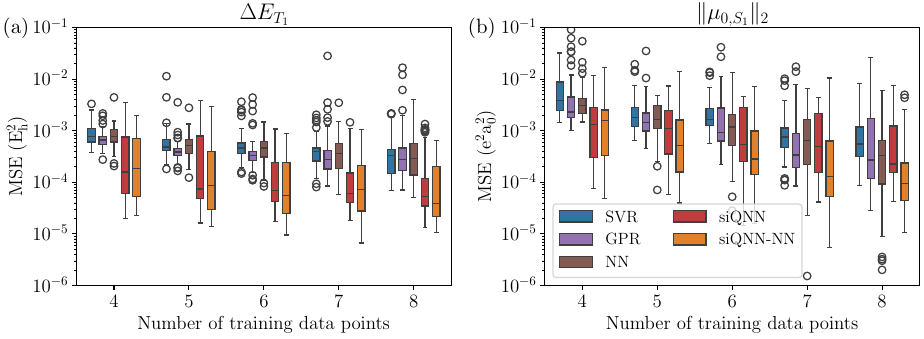}
\caption{\label{fig:results_BO-6-31G_BO_TD_example} Box plots of the MSE on the test datasets for 50 randomly sampled datasets $\mathcal{D}^L_l$ for the given models and training dataset sizes $L$ for LiH. (a) shows results for predicting the first transition energy $\Delta E_{T_1}$ and (b) the same for the norm of the first non-zero TDM $\mu_{0,S_1}$.}
\end{figure*}

\subsection{\label{sec:results_benchmark}Benchmark}
To draw conclusions about the performance of the models based on training dataset size independent of the specific choice of points, we sample multiple training datasets $\mathcal{D}_l^L\subset\mathcal{D}$. Each set $\mathcal{D}_l^L$ contains $L$ data points. For each $L$, we sample 50 independent datasets, i.e., $l=1,\dots,50$. As a test set, we choose the remaining data points for each $l$ and $L$, i.e., $\mathcal{D}\setminus\mathcal{D}_l^L$. In the following, all MSE scores shown are derived from the test dataset. We aim to simulate a realistic scenario where sampling is expensive, and the approximate locations of three different regions (attractive, equilibrium, and repulsive) along the dissociation curve are known. From each of these regions one data point is randomly chosen to account for the different behaviors in these regions. These three data points are then used as initial points for a Bayesian optimization, which determines the final data points in the training dataset $\mathcal{D}_l^L$. However, using a less informed sampling method, e.g., equally spaced training data points from random subsets of the given dataset yields similar results.

\subsubsection{\label{sec:results_lythiumhydrid}LiH}
For LiH we train the models on seven transition energies and three TDMs, see Figure~\ref{fig:method_target_functions} (a) and (b). We consider interatomic distances $R$ between $0.5$--$\qty{3.3}{\angstrom}$.

Figure~\ref{fig:results_BO-6-31G-LiH_best_example} shows two examples where the inference of the siQNN-NN along with three classical models is shown. The models have been trained on datasets of size (a) $L=4$ with the transition energy $\Delta E_{T_2}$ and (b) $L=6$ with the TDM $\|\boldsymbol{\mu_}{0,S_2}\|_2$. In (a) one can observe that all models besides the siQNN-NN agree on predicting a similar and more intuitive shape, given the training data points. The siQNN-NN has a less intuitive fitting bias, that matches the target function significantly better with an MSE score more than two orders of magnitude smaller than that of the other models. In (b), one can observe a similar behavior of the models, except that the kernel methods (SVR, GPR) exhibit overfitting. Here, the siQNN-NN fit has an MSE more than four times lower than the next best classical model. These observations are consistent with the results of the subsequent systematic study.

Figure~\ref{fig:results_BO-6-31G_BO_TD_example}~(a) exemplarily shows a boxplot of the MSE for various models for the first transition energy $\Delta E_{T_1}$. The median MSE on the test set for the quantum models  (siQNN and siQNN-NN) are similar to each other and nearly an order of magnitude lower than those of the classical models (SVR, GPR, NN), which themselves perform similarly across all training dataset sizes considered. Additionally, the variation of the MSE for the siQNN and siQNN-NN is significantly larger than that of the classical models. This is because, for some training dataset samples, the training points poorly represent the overall shape of the target function, causing all models to perform equally poorly. For other training data point combinations, the siQNN and siQNN-NN outperform the classical models by about an order of magnitude in MSE. In cases where other transition energies than $\Delta E_{T_1}$ are chosen, the performance of the models is similar (shown in Appendix~\ref{sec:appendix_mse_plots}).

Figure~\ref{fig:results_BO-6-31G_BO_TD_example}~(b) shows the MSE for the first TDM $\|\boldsymbol{\mu_}{0,S_1}\|_2$. Here, the overall discrepancy between quantum and classical models is smaller than for $\Delta E_{T_1}$. Nevertheless, for larger training dataset sizes, the siQNN-NN still performs about half an order of magnitude better than the classical models, but the performance of the siQNN is similar to that of the classical models. For the other TDMs (shown in Appendix~\ref{sec:appendix_mse_plots}), the siQNN and siQNN-NN perform similarly or worse than the classical models for small training dataset sizes; however, as the training dataset size increases, the siQNN-NN significantly outperforms the other models. This indicates that the siQNN alone is not expressive enough to account for the $R$ dependency and potentially to find the exact eigenbasis. This can be alleviated by parameterizing the interatomic distance with  a classical NN.

\begin{figure*}[t]
\includegraphics[width=1.0\textwidth,clip, trim= 0cm 0cm 0cm 0cm]{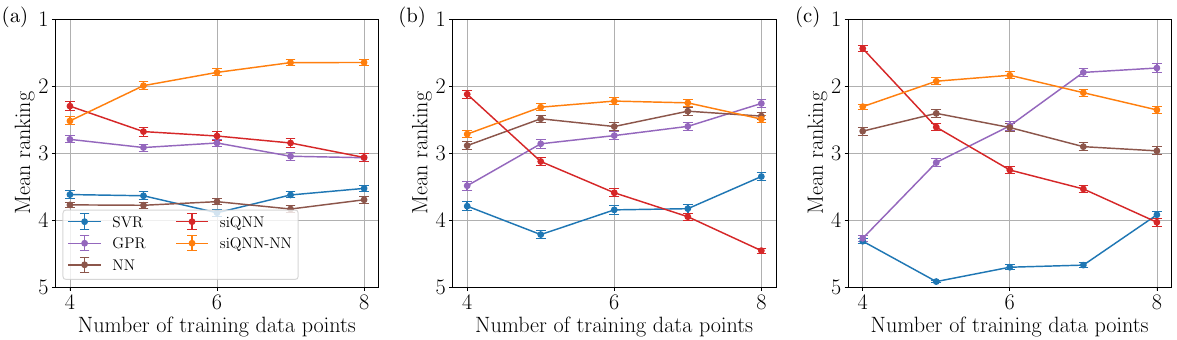}
\caption{\label{fig:results_combined_rankings_plot} Average rank of the models, broken down by training dataset size for (a) LiH, (b) $\mathrm{H_2}$, (c) and rectangular $\mathrm{H_4}$. Each point represents the relative rank of the respective model compared to all other models averaged over all considered targets and training samples for a molecule. The error bars indicate the standard errors of the mean values of those mean rankings.}
\end{figure*}

To summarize the performance across the different target functions for LiH, Figure~\ref{fig:results_combined_rankings_plot}~(a) shows the average rank of the models by MSE on these ten target functions for different training dataset sizes, with the smallest MSE ranked first. The best possible rank is 1.0, indicating that the model is the best for all target functions and training dataset samples, while a ranking of 5.0 represents the worst possible outcome. One can observe, that for the case $L=4$, the siQNN performs best, followed by the siQNN-NN. For all other training dataset sizes, the siQNN-NN is the best-performing model, achieving peak performance at a training dataset size of $L=8$, with a mean ranking of $1.646 \pm 0.045$, where the second-place ranking is nearly double that value. The ranking of the siQNN declines with increasing training dataset size. This can be explained because the MSE score for the siQNN remains roughly constant while the performance of the other models improves with increasing training dataset size. This indicates that, in most cases, the siQNN finds a good approximation of the target function but is not expressive enough to find a set of parameters that fits the targets for all values of $R$. Among the classical models, the GPR performs the best, while the NN and SVR perform similarly.

\subsubsection{$\mathrm{H}_2$}
For $\mathrm{H}_2$ we train the models on seven transition energies and two TDMs, see Figure~\ref{fig:method_target_functions} (c) and (d). We consider interatomic distances $R$ between $0.4$--$\qty{2.6}{\angstrom}$. Figure~\ref{fig:results_combined_rankings_plot}~(b) shows the ranking plot for the performance on $\mathrm{H}_2$. As for LiH, one of the ground-state based models (siQNN or siQNN-NN) is predominantly the best-performing model, with the only exception occurring at eight training data points, where the GPR outperforms the siQNN-NN, and the NN performs within the error range of the siQNN-NN. This is likely due to the fact that several target functions for $\mathrm{H}_2$ are relatively trivial. Furthermore, the performance of the NN in the case of $\mathrm{H}_2$, with an average rank of about 2.5, is significantly better than in the case of LiH. This indicates that lower complexity of the molecule is more compatible with the small NN.

Some of the target functions (Figure~\ref{fig:method_target_functions}), such as $\Delta E_{T_{1}}$, are relatively trivial, which is also reflected in the MSE scores of the models (shown in Appendix~\ref{sec:appendix_mse_plots}). For these trivial target functions, the MSE for some classical models is already in the order of $10^{-5}$ with a training dataset size of four, and it decreases with increasing training dataset size down to about $10^{-7}$. The quantum-enhanced models perform about an order of magnitude worse than the classical models for these functions because the increased feature space adds additional complexity to the learning task, which introduces unnecessary complexity and therefore overfitting. Further, as explained in Section~\ref{sec:methods_training}, we stop the training of the (Q)NN-based models when $\mathcal{L}_{\rm t} = 10^{-6}$ is reached. Therefore, if the test MSE of these models is close to or below $10^{-6}$, further training with a smaller $\mathcal{L}_{\rm t}$ could reduce the test MSE even further for these models, potentially aligning them with models that do not use a $\mathcal{L}_{\rm t}$ such as the kernel methods, i.e., SVR or GPR. Consequently, for target functions and training dataset sizes with MSE values close to or below $\mathcal{L}_{\rm t}$, comparisons between the (Q)NNs and the kernel methods are of limited significance. This effect becomes more relevant for increasing training dataset size.

For more complex target functions, such as $\Delta E_{T_{2}}$, where the test MSE score of the classical models is mostly above $10^{-5}$ for several training dataset sizes, the siQNN-NN outperforms the NN model and the other classical models for small $L$. For some target functions, kernel methods exhibit  MSE scores that are orders of magnitude larger than the MSE score of the NN. This is mostly due to overfitting which is difficult to prevent in the low-data regime. 

\subsubsection{$\mathrm{H}_4$}
The target functions for $\mathrm{H}_{4}$ are shown in Figure~\ref{fig:method_target_functions} (e) and (f). Three transition energies are presented over the region $0.5$--$\qty{2.0}{\angstrom}$ for $R$ and the distance between the Hydrogen atoms on the other axis of the rectangular $\mathrm{H}_4$ is equal to $\qty{2.0}{\angstrom}$ for all target functions. Additionally, three TDMs are shown in reduced regions beyond an avoided crossing ($0.8$--$\qty{2.0}{\angstrom}$ and $1.0$--$\qty{2.0}{\angstrom}$). Furthermore, one transition energy is considered in the reduced region beyond the avoided crossing ($1.0$ and $\qty{2.0}{\angstrom}$) to at least include the first two singlet and triplet energy states. The corresponding dataset sizes are reduced to the data points within the reduced regions, i.e., $M=80$ and $M=65$, respectively.

The ranking plot for $\mathrm{H}_4$ is shown in Figure~\ref{fig:results_combined_rankings_plot}~(c). It is similar to the one for $\mathrm{H}_2$, but the SVR performs worse overall, while the GPR performs better, already being the best model for seven training data points. The siQNN-NN performs significantly better than the NN for all training dataset sizes with a peak performance of $1.837 \pm 0.053$ for six training data points, while the siQNN has a peak performance of $1.437 \pm 0.038$ for four training data points.

Overall, one can observe that the classical models outperform the quantum-enhanced models for relatively trivial target functions already for the considered low data regime. However, for more complex target functions, the quantum-enhanced models outperform the classical models, as the additional information from the ground state can be leveraged to achieve better prediction performances.

\section{discussion}
\label{sec:discussion}
In this work, we investigated the effectiveness of enhancing the performance of a machine learning model trained to predict excited-state properties of a molecule in different geometric configurations. We achieved this by training a QNN directly on the molecular ground state to extract information from the wave function. The proposed model consists of a symmetry-respecting QNN combined with an NN, with the observable measured on the quantum state consisting only of commuting Pauli-strings. \sout{Pauli-$Z$ operators.} The performance of the siQNN-NN and siQNN were significantly better than the performance of the models learning only on classical data for small training dataset sizes across all molecules with a test MSE reduction of up to two orders of magnitude.

From these observations, we deduce that the siQNN effectively extracts valuable information from the ground state and provides a good estimate of various target functions in this study. For very small training datasets, the additional complexity of the siQNN-NN leads to overfitting which decreases the performance relative to the siQNN. As the training dataset size increases, the siQNN, however, struggles to describe the training data points to the given accuracy, and the additional degrees of freedom in the NN become valuable to account for the $R$ dependence and the potentially imperfect preparation of the eigenbasis of the target observable with the siQNN. With increasing data set size, the performance gap between classical and quantum models closes. We observe that the exact number of data points where this can be observed is dependent on the complexity of the target function, that is, more complex target functions require more data points for classical models to be competitive with quantum models.

We note several limitations of our study. Firstly, we expect that for each target function and molecule there exist a number of training data points (depending on its complexity) above which the models learning purely on the classical data reach sufficient test dataset accuracy on average and therefore the additional effort of preparing the quantum information is not reasonable anymore. This is because for all data points, the ground state needs to be provided, introducing a relevant overhead relative to models learning only on the classical data. Further, to enable our model to predict excited-state properties, training values of those properties need to be derived using another technique. Therefore, our model is relevant for molecules where the properties of interest for different geometric configurations are calculable but with significant effort, while the ground state is less computationally demanding. Given that excited-state properties are generally more complex than ground-state properties, there are several molecules and properties that meet this condition.

Secondly, the substructure of the siQNN has similarities to the structure of a quantum convolutional neural network, which is known to be simulatable. This suggests the possibility that the siQNN is efficiently classically simulatable. However, the envisioned use case for the procedure described in this work is a scenario where molecular ground state simulations are performed on a quantum computer. Adding a comparatively small QNN as suggested here is then a relatively small overhead to obtain excited state properties on top of ground state properties. Still, it could be an interesting further research path to construct a purely classical model inspired by our quantum-enhanced model that could extract valuable information and investigate whether it can do this even more efficiently from the ground, for example, using classical shadows~\cite{classical_shadows}. 

Finally, this study was conducted using exact ground states. Investigating the effect of imperfect simulations of molecular ground states as input is an interesting research question. In Appendix~\ref{sec:appendix_shot_noise_influence} we further estimate the effect of shot noise on the performance of our machine learning approach and find promising results. Regarding decoherence and gate noise, given that using the circuit in the paper extends the circuit depth only logarithmically in the number of qubits compared to the preparation of the ground state, we have good reason to believe that our method does not significantly deviate from results obtained from NISQ state-preparation circuits (e.g. VQE). More specifically, the results will be comparable to other types of QNNs of similar size (see e.g. Ref.~\cite{Kreplin2024reductionoffinite}, which has used a similar observable on real quantum computers). A more rigorous investigation into the effect of shot noise and additionally the effect of other noise types apparent in current quantum computers could be conducted in follow-up work.

In order to estimate the scaling of our model to larger molecules, we confirmed its applicability and performance advantages with different system sizes, namely four, five, and six orbitals. However, these are small problem sizes compared to many molecules of interest, such as iron complexes, which require too many qubits to be prepared on today's quantum simulators and devices. A potential problem for these larger system sizes could be the complexity of finding the eigenbasis of the target observable. A potential approach could be to prepare multiple copies of the ground state followed by QNNs, whose measurement is added up and, therefore, optimized in parallel, so that the different copies could prepare different common eigenstates of commuting subsets of the target observable, to break down the eigenbasis search task.

In conclusion, we propose a procedure to derive excited-state properties in an efficient way. We consider this approach valuable to analyzing molecular properties and understanding and predicting their behaviors. We view this study as a starting point for further investigations into extracting information from the ground state for excited-state properties.

\begin{acknowledgments}
This work was supported by the German Federal Ministry
of Economic Affairs and Climate Action through the
projects AQUAS (grant no. 01MQ22003D). The authors would like
to thank Jan Schnabel and David A. Kreplin for fruitful discussions. The authors disclose the use of LLM-based tools for grammar and spell-checking.
\end{acknowledgments}

\balance

\bibliography{apssamp}

\begin{thebibliography}{39}%
\makeatletter
\providecommand \@ifxundefined [1]{%
 \@ifx{#1\undefined}
}%
\providecommand \@ifnum [1]{%
 \ifnum #1\expandafter \@firstoftwo
 \else \expandafter \@secondoftwo
 \fi
}%
\providecommand \@ifx [1]{%
 \ifx #1\expandafter \@firstoftwo
 \else \expandafter \@secondoftwo
 \fi
}%
\providecommand \natexlab [1]{#1}%
\providecommand \enquote  [1]{``#1''}%
\providecommand \bibnamefont  [1]{#1}%
\providecommand \bibfnamefont [1]{#1}%
\providecommand \citenamefont [1]{#1}%
\providecommand \href@noop [0]{\@secondoftwo}%
\providecommand \href [0]{\begingroup \@sanitize@url \@href}%
\providecommand \@href[1]{\@@startlink{#1}\@@href}%
\providecommand \@@href[1]{\endgroup#1\@@endlink}%
\providecommand \@sanitize@url [0]{\catcode `\\12\catcode `\$12\catcode `\&12\catcode `\#12\catcode `\^12\catcode `\_12\catcode `\%12\relax}%
\providecommand \@@startlink[1]{}%
\providecommand \@@endlink[0]{}%
\providecommand \url  [0]{\begingroup\@sanitize@url \@url }%
\providecommand \@url [1]{\endgroup\@href {#1}{\urlprefix }}%
\providecommand \urlprefix  [0]{URL }%
\providecommand \Eprint [0]{\href }%
\providecommand \doibase [0]{https://doi.org/}%
\providecommand \selectlanguage [0]{\@gobble}%
\providecommand \bibinfo  [0]{\@secondoftwo}%
\providecommand \bibfield  [0]{\@secondoftwo}%
\providecommand \translation [1]{[#1]}%
\providecommand \BibitemOpen [0]{}%
\providecommand \bibitemStop [0]{}%
\providecommand \bibitemNoStop [0]{.\EOS\space}%
\providecommand \EOS [0]{\spacefactor3000\relax}%
\providecommand \BibitemShut  [1]{\csname bibitem#1\endcsname}%
\let\auto@bib@innerbib\@empty
\bibitem [{\citenamefont {Serrano-Andrés}\ and\ \citenamefont {Merchán}(2005)}]{ExcitedStateOverview2005}%
  \BibitemOpen
  \bibfield  {author} {\bibinfo {author} {\bibfnamefont {L.}~\bibnamefont {Serrano-Andrés}}\ and\ \bibinfo {author} {\bibfnamefont {M.}~\bibnamefont {Merchán}},\ }\bibfield  {title} {\bibinfo {title} {Quantum chemistry of the excited state: 2005 overview},\ }\href {https://doi.org/https://doi.org/10.1016/j.theochem.2005.03.020} {\bibfield  {journal} {\bibinfo  {journal} {Journal of Molecular Structure: THEOCHEM}\ }\textbf {\bibinfo {volume} {729}},\ \bibinfo {pages} {99} (\bibinfo {year} {2005})}\BibitemShut {NoStop}%
\bibitem [{\citenamefont {Shahnavi}\ \emph {et~al.}(2016)\citenamefont {Shahnavi}, \citenamefont {Ahmed}, \citenamefont {Anwar}, \citenamefont {Sheraz},\ and\ \citenamefont {Sikorski}}]{PhotostabilityOnDrugs}%
  \BibitemOpen
  \bibfield  {author} {\bibinfo {author} {\bibfnamefont {I.}~\bibnamefont {Shahnavi}}, \bibinfo {author} {\bibfnamefont {S.}~\bibnamefont {Ahmed}}, \bibinfo {author} {\bibfnamefont {Z.}~\bibnamefont {Anwar}}, \bibinfo {author} {\bibfnamefont {M.}~\bibnamefont {Sheraz}},\ and\ \bibinfo {author} {\bibfnamefont {M.}~\bibnamefont {Sikorski}},\ }\bibfield  {title} {\bibinfo {title} {Photostability and photostabilization of drugs and drug products},\ }\href {https://doi.org/10.1155/2016/8135608} {\bibfield  {journal} {\bibinfo  {journal} {International Journal of Photoenergy}\ }\textbf {\bibinfo {volume} {2016}},\ \bibinfo {pages} {1} (\bibinfo {year} {2016})}\BibitemShut {NoStop}%
\bibitem [{\citenamefont {Vöhringer-Martinez}\ and\ \citenamefont {Toro-Labbé}(2012)}]{Voehringer2012}%
  \BibitemOpen
  \bibfield  {author} {\bibinfo {author} {\bibfnamefont {E.}~\bibnamefont {Vöhringer-Martinez}}\ and\ \bibinfo {author} {\bibfnamefont {A.}~\bibnamefont {Toro-Labbé}},\ }\bibfield  {title} {\bibinfo {title} {Understanding the physics and chemistry of reaction mechanisms from atomic contributions: A reaction force perspective},\ }\href {https://doi.org/10.1021/jp303075k} {\bibfield  {journal} {\bibinfo  {journal} {The Journal of Physical Chemistry A}\ }\textbf {\bibinfo {volume} {116}},\ \bibinfo {pages} {7419} (\bibinfo {year} {2012})}\BibitemShut {NoStop}%
\bibitem [{\citenamefont {Cao}\ \emph {et~al.}(2018)\citenamefont {Cao}, \citenamefont {Romero},\ and\ \citenamefont {Aspuru-Guzik}}]{QC_drug_discovery}%
  \BibitemOpen
  \bibfield  {author} {\bibinfo {author} {\bibfnamefont {Y.}~\bibnamefont {Cao}}, \bibinfo {author} {\bibfnamefont {J.}~\bibnamefont {Romero}},\ and\ \bibinfo {author} {\bibfnamefont {A.}~\bibnamefont {Aspuru-Guzik}},\ }\bibfield  {title} {\bibinfo {title} {Potential of quantum computing for drug discovery},\ }\href {https://doi.org/10.1147/JRD.2018.2888987} {\bibfield  {journal} {\bibinfo  {journal} {IBM Journal of Research and Development}\ }\textbf {\bibinfo {volume} {62}},\ \bibinfo {pages} {6:1} (\bibinfo {year} {2018})}\BibitemShut {NoStop}%
\bibitem [{\citenamefont {von Burg}\ \emph {et~al.}(2021)\citenamefont {von Burg}, \citenamefont {Low}, \citenamefont {H\"aner}, \citenamefont {Steiger}, \citenamefont {Reiher}, \citenamefont {Roetteler},\ and\ \citenamefont {Troyer}}]{QC_catalyst_discovery}%
  \BibitemOpen
  \bibfield  {author} {\bibinfo {author} {\bibfnamefont {V.}~\bibnamefont {von Burg}}, \bibinfo {author} {\bibfnamefont {G.~H.}\ \bibnamefont {Low}}, \bibinfo {author} {\bibfnamefont {T.}~\bibnamefont {H\"aner}}, \bibinfo {author} {\bibfnamefont {D.~S.}\ \bibnamefont {Steiger}}, \bibinfo {author} {\bibfnamefont {M.}~\bibnamefont {Reiher}}, \bibinfo {author} {\bibfnamefont {M.}~\bibnamefont {Roetteler}},\ and\ \bibinfo {author} {\bibfnamefont {M.}~\bibnamefont {Troyer}},\ }\bibfield  {title} {\bibinfo {title} {Quantum computing enhanced computational catalysis},\ }\href {https://doi.org/10.1103/PhysRevResearch.3.033055} {\bibfield  {journal} {\bibinfo  {journal} {Phys. Rev. Res.}\ }\textbf {\bibinfo {volume} {3}},\ \bibinfo {pages} {033055} (\bibinfo {year} {2021})}\BibitemShut {NoStop}%
\bibitem [{\citenamefont {Knowles}\ and\ \citenamefont {Handy}(1984)}]{FCI}%
  \BibitemOpen
  \bibfield  {author} {\bibinfo {author} {\bibfnamefont {P.}~\bibnamefont {Knowles}}\ and\ \bibinfo {author} {\bibfnamefont {N.}~\bibnamefont {Handy}},\ }\bibfield  {title} {\bibinfo {title} {A new determinant-based full configuration interaction method},\ }\href {https://doi.org/https://doi.org/10.1016/0009-2614(84)85513-X} {\bibfield  {journal} {\bibinfo  {journal} {Chemical Physics Letters}\ }\textbf {\bibinfo {volume} {111}},\ \bibinfo {pages} {315} (\bibinfo {year} {1984})}\BibitemShut {NoStop}%
\bibitem [{\citenamefont {Sun}\ \emph {et~al.}(2017{\natexlab{a}})\citenamefont {Sun}, \citenamefont {Yang},\ and\ \citenamefont {Chan}}]{CASSCF}%
  \BibitemOpen
  \bibfield  {author} {\bibinfo {author} {\bibfnamefont {Q.}~\bibnamefont {Sun}}, \bibinfo {author} {\bibfnamefont {J.}~\bibnamefont {Yang}},\ and\ \bibinfo {author} {\bibfnamefont {G.~K.-L.}\ \bibnamefont {Chan}},\ }\bibfield  {title} {\bibinfo {title} {A general second order complete active space self-consistent-field solver for large-scale systems},\ }\href {https://doi.org/10.1016/j.cplett.2017.03.004} {\bibfield  {journal} {\bibinfo  {journal} {Chemical Physics Letters}\ }\textbf {\bibinfo {volume} {683}},\ \bibinfo {pages} {291–299} (\bibinfo {year} {2017}{\natexlab{a}})}\BibitemShut {NoStop}%
\bibitem [{\citenamefont {Santagati}\ \emph {et~al.}(2024)\citenamefont {Santagati}, \citenamefont {Aspuru-Guzik}, \citenamefont {Babbush}, \citenamefont {Degroote}, \citenamefont {González}, \citenamefont {Kyoseva}, \citenamefont {Moll}, \citenamefont {Oppel}, \citenamefont {Parrish}, \citenamefont {Rubin}, \citenamefont {Streif}, \citenamefont {Tautermann}, \citenamefont {Weiss}, \citenamefont {Wiebe},\ and\ \citenamefont {Utschig-Utschig}}]{DrugDesignOnQC}%
  \BibitemOpen
  \bibfield  {author} {\bibinfo {author} {\bibfnamefont {R.}~\bibnamefont {Santagati}}, \bibinfo {author} {\bibfnamefont {A.}~\bibnamefont {Aspuru-Guzik}}, \bibinfo {author} {\bibfnamefont {R.}~\bibnamefont {Babbush}}, \bibinfo {author} {\bibfnamefont {M.}~\bibnamefont {Degroote}}, \bibinfo {author} {\bibfnamefont {L.}~\bibnamefont {González}}, \bibinfo {author} {\bibfnamefont {E.}~\bibnamefont {Kyoseva}}, \bibinfo {author} {\bibfnamefont {N.}~\bibnamefont {Moll}}, \bibinfo {author} {\bibfnamefont {M.}~\bibnamefont {Oppel}}, \bibinfo {author} {\bibfnamefont {R.~M.}\ \bibnamefont {Parrish}}, \bibinfo {author} {\bibfnamefont {N.~C.}\ \bibnamefont {Rubin}}, \bibinfo {author} {\bibfnamefont {M.}~\bibnamefont {Streif}}, \bibinfo {author} {\bibfnamefont {C.~S.}\ \bibnamefont {Tautermann}}, \bibinfo {author} {\bibfnamefont {H.}~\bibnamefont {Weiss}}, \bibinfo {author} {\bibfnamefont {N.}~\bibnamefont {Wiebe}},\ and\ \bibinfo {author} {\bibfnamefont {C.}~\bibnamefont {Utschig-Utschig}},\ }\bibfield  {title}
  {\bibinfo {title} {Drug design on quantum computers},\ }\href {https://doi.org/10.1038/s41567-024-02411-5} {\bibfield  {journal} {\bibinfo  {journal} {Nature Physics}\ }\textbf {\bibinfo {volume} {20}},\ \bibinfo {pages} {549–557} (\bibinfo {year} {2024})}\BibitemShut {NoStop}%
\bibitem [{\citenamefont {Tilly}\ \emph {et~al.}(2022)\citenamefont {Tilly}, \citenamefont {Chen}, \citenamefont {Cao}, \citenamefont {Picozzi}, \citenamefont {Setia}, \citenamefont {Li}, \citenamefont {Grant}, \citenamefont {Wossnig}, \citenamefont {Rungger}, \citenamefont {Booth},\ and\ \citenamefont {Tennyson}}]{VQE}%
  \BibitemOpen
  \bibfield  {author} {\bibinfo {author} {\bibfnamefont {J.}~\bibnamefont {Tilly}}, \bibinfo {author} {\bibfnamefont {H.}~\bibnamefont {Chen}}, \bibinfo {author} {\bibfnamefont {S.}~\bibnamefont {Cao}}, \bibinfo {author} {\bibfnamefont {D.}~\bibnamefont {Picozzi}}, \bibinfo {author} {\bibfnamefont {K.}~\bibnamefont {Setia}}, \bibinfo {author} {\bibfnamefont {Y.}~\bibnamefont {Li}}, \bibinfo {author} {\bibfnamefont {E.}~\bibnamefont {Grant}}, \bibinfo {author} {\bibfnamefont {L.}~\bibnamefont {Wossnig}}, \bibinfo {author} {\bibfnamefont {I.}~\bibnamefont {Rungger}}, \bibinfo {author} {\bibfnamefont {G.~H.}\ \bibnamefont {Booth}},\ and\ \bibinfo {author} {\bibfnamefont {J.}~\bibnamefont {Tennyson}},\ }\bibfield  {title} {\bibinfo {title} {The variational quantum eigensolver: A review of methods and best practices},\ }\href {https://doi.org/https://doi.org/10.1016/j.physrep.2022.08.003} {\bibfield  {journal} {\bibinfo  {journal} {Physics Reports}\ }\textbf {\bibinfo {volume} {986}},\ \bibinfo {pages} {1}
  (\bibinfo {year} {2022})}\BibitemShut {NoStop}%
\bibitem [{\citenamefont {Cleve}\ \emph {et~al.}(1998)\citenamefont {Cleve}, \citenamefont {Ekert}, \citenamefont {Macchiavello},\ and\ \citenamefont {Mosca}}]{QPE}%
  \BibitemOpen
  \bibfield  {author} {\bibinfo {author} {\bibfnamefont {R.}~\bibnamefont {Cleve}}, \bibinfo {author} {\bibfnamefont {A.}~\bibnamefont {Ekert}}, \bibinfo {author} {\bibfnamefont {C.}~\bibnamefont {Macchiavello}},\ and\ \bibinfo {author} {\bibfnamefont {M.}~\bibnamefont {Mosca}},\ }\bibfield  {title} {\bibinfo {title} {Quantum algorithms revisited},\ }\href {https://doi.org/10.1098/rspa.1998.0164} {\bibfield  {journal} {\bibinfo  {journal} {Proceedings of the Royal Society of London. Series A: Mathematical, Physical and Engineering Sciences}\ }\textbf {\bibinfo {volume} {454}},\ \bibinfo {pages} {339–354} (\bibinfo {year} {1998})}\BibitemShut {NoStop}%
\bibitem [{\citenamefont {Córcoles}\ \emph {et~al.}(2020)\citenamefont {Córcoles}, \citenamefont {Kandala}, \citenamefont {Javadi-Abhari}, \citenamefont {McClure}, \citenamefont {Cross}, \citenamefont {Temme}, \citenamefont {Nation}, \citenamefont {Steffen},\ and\ \citenamefont {Gambetta}}]{NISQ}%
  \BibitemOpen
  \bibfield  {author} {\bibinfo {author} {\bibfnamefont {A.~D.}\ \bibnamefont {Córcoles}}, \bibinfo {author} {\bibfnamefont {A.}~\bibnamefont {Kandala}}, \bibinfo {author} {\bibfnamefont {A.}~\bibnamefont {Javadi-Abhari}}, \bibinfo {author} {\bibfnamefont {D.~T.}\ \bibnamefont {McClure}}, \bibinfo {author} {\bibfnamefont {A.~W.}\ \bibnamefont {Cross}}, \bibinfo {author} {\bibfnamefont {K.}~\bibnamefont {Temme}}, \bibinfo {author} {\bibfnamefont {P.~D.}\ \bibnamefont {Nation}}, \bibinfo {author} {\bibfnamefont {M.}~\bibnamefont {Steffen}},\ and\ \bibinfo {author} {\bibfnamefont {J.~M.}\ \bibnamefont {Gambetta}},\ }\bibfield  {title} {\bibinfo {title} {Challenges and opportunities of near-term quantum computing systems},\ }\href {https://doi.org/10.1109/JPROC.2019.2954005} {\bibfield  {journal} {\bibinfo  {journal} {Proceedings of the IEEE}\ }\textbf {\bibinfo {volume} {108}},\ \bibinfo {pages} {1338} (\bibinfo {year} {2020})}\BibitemShut {NoStop}%
\bibitem [{\citenamefont {Lee}\ \emph {et~al.}(2023)\citenamefont {Lee}, \citenamefont {Lee}, \citenamefont {Zhai}, \citenamefont {Tong}, \citenamefont {Dalzell}, \citenamefont {Kumar}, \citenamefont {Helms}, \citenamefont {Gray}, \citenamefont {Cui}, \citenamefont {Liu}, \citenamefont {Kastoryano}, \citenamefont {Babbush}, \citenamefont {Preskill}, \citenamefont {Reichman}, \citenamefont {Campbell}, \citenamefont {Valeev}, \citenamefont {Lin},\ and\ \citenamefont {Chan}}]{QPEHighOverlap}%
  \BibitemOpen
  \bibfield  {author} {\bibinfo {author} {\bibfnamefont {S.}~\bibnamefont {Lee}}, \bibinfo {author} {\bibfnamefont {J.}~\bibnamefont {Lee}}, \bibinfo {author} {\bibfnamefont {H.}~\bibnamefont {Zhai}}, \bibinfo {author} {\bibfnamefont {Y.}~\bibnamefont {Tong}}, \bibinfo {author} {\bibfnamefont {A.}~\bibnamefont {Dalzell}}, \bibinfo {author} {\bibfnamefont {A.}~\bibnamefont {Kumar}}, \bibinfo {author} {\bibfnamefont {P.}~\bibnamefont {Helms}}, \bibinfo {author} {\bibfnamefont {J.}~\bibnamefont {Gray}}, \bibinfo {author} {\bibfnamefont {Z.-H.}\ \bibnamefont {Cui}}, \bibinfo {author} {\bibfnamefont {W.}~\bibnamefont {Liu}}, \bibinfo {author} {\bibfnamefont {M.}~\bibnamefont {Kastoryano}}, \bibinfo {author} {\bibfnamefont {R.}~\bibnamefont {Babbush}}, \bibinfo {author} {\bibfnamefont {J.}~\bibnamefont {Preskill}}, \bibinfo {author} {\bibfnamefont {D.}~\bibnamefont {Reichman}}, \bibinfo {author} {\bibfnamefont {E.}~\bibnamefont {Campbell}}, \bibinfo {author} {\bibfnamefont {E.}~\bibnamefont {Valeev}}, \bibinfo
  {author} {\bibfnamefont {L.}~\bibnamefont {Lin}},\ and\ \bibinfo {author} {\bibfnamefont {G.}~\bibnamefont {Chan}},\ }\bibfield  {title} {\bibinfo {title} {Evaluating the evidence for exponential quantum advantage in ground-state quantum chemistry},\ }\href {https://doi.org/10.1038/s41467-023-37587-6} {\bibfield  {journal} {\bibinfo  {journal} {Nature Communications}\ }\textbf {\bibinfo {volume} {14}} (\bibinfo {year} {2023})}\BibitemShut {NoStop}%
\bibitem [{\citenamefont {Ceroni}\ \emph {et~al.}(2023)\citenamefont {Ceroni}, \citenamefont {Stetina}, \citenamefont {Kieferova}, \citenamefont {Marrero}, \citenamefont {Arrazola},\ and\ \citenamefont {Wiebe}}]{ceroni2023generatingapproximategroundstates}%
  \BibitemOpen
  \bibfield  {author} {\bibinfo {author} {\bibfnamefont {J.}~\bibnamefont {Ceroni}}, \bibinfo {author} {\bibfnamefont {T.~F.}\ \bibnamefont {Stetina}}, \bibinfo {author} {\bibfnamefont {M.}~\bibnamefont {Kieferova}}, \bibinfo {author} {\bibfnamefont {C.~O.}\ \bibnamefont {Marrero}}, \bibinfo {author} {\bibfnamefont {J.~M.}\ \bibnamefont {Arrazola}},\ and\ \bibinfo {author} {\bibfnamefont {N.}~\bibnamefont {Wiebe}},\ }\href@noop {} {\bibinfo {title} {\href{https://arxiv.org/abs/2210.05489}{Generating Approximate Ground States of Molecules Using Quantum Machine Learning}}} (\bibinfo {year} {2023})\BibitemShut {NoStop}%
\bibitem [{\citenamefont {Robledo-Moreno}\ \emph {et~al.}(2024)\citenamefont {Robledo-Moreno}, \citenamefont {Motta}, \citenamefont {Haas}, \citenamefont {Javadi-Abhari}, \citenamefont {Jurcevic}, \citenamefont {Kirby}, \citenamefont {Martiel}, \citenamefont {Sharma}, \citenamefont {Sharma}, \citenamefont {Shirakawa}, \citenamefont {Sitdikov}, \citenamefont {Sun}, \citenamefont {Sung}, \citenamefont {Takita}, \citenamefont {Tran}, \citenamefont {Yunoki},\ and\ \citenamefont {Mezzacapo}}]{robledomoreno2024chemistryexactsolutionsquantumcentric}%
  \BibitemOpen
  \bibfield  {author} {\bibinfo {author} {\bibfnamefont {J.}~\bibnamefont {Robledo-Moreno}}, \bibinfo {author} {\bibfnamefont {M.}~\bibnamefont {Motta}}, \bibinfo {author} {\bibfnamefont {H.}~\bibnamefont {Haas}}, \bibinfo {author} {\bibfnamefont {A.}~\bibnamefont {Javadi-Abhari}}, \bibinfo {author} {\bibfnamefont {P.}~\bibnamefont {Jurcevic}}, \bibinfo {author} {\bibfnamefont {W.}~\bibnamefont {Kirby}}, \bibinfo {author} {\bibfnamefont {S.}~\bibnamefont {Martiel}}, \bibinfo {author} {\bibfnamefont {K.}~\bibnamefont {Sharma}}, \bibinfo {author} {\bibfnamefont {S.}~\bibnamefont {Sharma}}, \bibinfo {author} {\bibfnamefont {T.}~\bibnamefont {Shirakawa}}, \bibinfo {author} {\bibfnamefont {I.}~\bibnamefont {Sitdikov}}, \bibinfo {author} {\bibfnamefont {R.-Y.}\ \bibnamefont {Sun}}, \bibinfo {author} {\bibfnamefont {K.~J.}\ \bibnamefont {Sung}}, \bibinfo {author} {\bibfnamefont {M.}~\bibnamefont {Takita}}, \bibinfo {author} {\bibfnamefont {M.~C.}\ \bibnamefont {Tran}}, \bibinfo {author} {\bibfnamefont
  {S.}~\bibnamefont {Yunoki}},\ and\ \bibinfo {author} {\bibfnamefont {A.}~\bibnamefont {Mezzacapo}},\ }\href@noop {} {\bibinfo {title} {\href{https://arxiv.org/abs/2405.05068}{Chemistry Beyond Exact Solutions on a Quantum-Centric Supercomputer}}} (\bibinfo {year} {2024})\BibitemShut {NoStop}%
\bibitem [{\citenamefont {{P. B. Armentrout}}(1991)}]{excited_states_configurations}%
  \BibitemOpen
  \bibfield  {author} {\bibinfo {author} {\bibnamefont {{P. B. Armentrout}}},\ }\bibfield  {title} {\bibinfo {title} {Chemistry of excited electronic states},\ }\href {https://doi.org/10.1126/science.251.4990.175} {\bibfield  {journal} {\bibinfo  {journal} {Science}\ }\textbf {\bibinfo {volume} {251}},\ \bibinfo {pages} {175} (\bibinfo {year} {1991})}\BibitemShut {NoStop}%
\bibitem [{\citenamefont {González}\ \emph {et~al.}(2012)\citenamefont {González}, \citenamefont {Escudero},\ and\ \citenamefont {Serrano-Andrés}}]{ExcitedStateChallenges}%
  \BibitemOpen
  \bibfield  {author} {\bibinfo {author} {\bibfnamefont {L.}~\bibnamefont {González}}, \bibinfo {author} {\bibfnamefont {D.}~\bibnamefont {Escudero}},\ and\ \bibinfo {author} {\bibfnamefont {L.}~\bibnamefont {Serrano-Andrés}},\ }\bibfield  {title} {\bibinfo {title} {Progress and challenges in the calculation of electronic excited states},\ }\href {https://doi.org/https://doi.org/10.1002/cphc.201100200} {\bibfield  {journal} {\bibinfo  {journal} {ChemPhysChem}\ }\textbf {\bibinfo {volume} {13}},\ \bibinfo {pages} {28} (\bibinfo {year} {2012})}\BibitemShut {NoStop}%
\bibitem [{\citenamefont {Higgott}\ \emph {et~al.}(2019)\citenamefont {Higgott}, \citenamefont {Wang},\ and\ \citenamefont {Brierley}}]{VQD}%
  \BibitemOpen
  \bibfield  {author} {\bibinfo {author} {\bibfnamefont {O.}~\bibnamefont {Higgott}}, \bibinfo {author} {\bibfnamefont {D.}~\bibnamefont {Wang}},\ and\ \bibinfo {author} {\bibfnamefont {S.}~\bibnamefont {Brierley}},\ }\bibfield  {title} {\bibinfo {title} {Variational quantum computation of excited states},\ }\href {https://doi.org/10.22331/q-2019-07-01-156} {\bibfield  {journal} {\bibinfo  {journal} {Quantum}\ }\textbf {\bibinfo {volume} {3}},\ \bibinfo {pages} {156} (\bibinfo {year} {2019})}\BibitemShut {NoStop}%
\bibitem [{\citenamefont {Nakanishi}\ \emph {et~al.}(2019)\citenamefont {Nakanishi}, \citenamefont {Mitarai},\ and\ \citenamefont {Fujii}}]{ssVQE}%
  \BibitemOpen
  \bibfield  {author} {\bibinfo {author} {\bibfnamefont {K.~M.}\ \bibnamefont {Nakanishi}}, \bibinfo {author} {\bibfnamefont {K.}~\bibnamefont {Mitarai}},\ and\ \bibinfo {author} {\bibfnamefont {K.}~\bibnamefont {Fujii}},\ }\bibfield  {title} {\bibinfo {title} {Subspace-search variational quantum eigensolver for excited states},\ }\href {https://doi.org/10.1103/PhysRevResearch.1.033062} {\bibfield  {journal} {\bibinfo  {journal} {Phys. Rev. Res.}\ }\textbf {\bibinfo {volume} {1}},\ \bibinfo {pages} {033062} (\bibinfo {year} {2019})}\BibitemShut {NoStop}%
\bibitem [{\citenamefont {Cerezo}\ \emph {et~al.}(2022)\citenamefont {Cerezo}, \citenamefont {Verdon}, \citenamefont {Huang}, \citenamefont {Cincio},\ and\ \citenamefont {Coles}}]{Cerezo2022}%
  \BibitemOpen
  \bibfield  {author} {\bibinfo {author} {\bibfnamefont {M.}~\bibnamefont {Cerezo}}, \bibinfo {author} {\bibfnamefont {G.}~\bibnamefont {Verdon}}, \bibinfo {author} {\bibfnamefont {H.-Y.}\ \bibnamefont {Huang}}, \bibinfo {author} {\bibfnamefont {L.}~\bibnamefont {Cincio}},\ and\ \bibinfo {author} {\bibfnamefont {P.~J.}\ \bibnamefont {Coles}},\ }\bibfield  {title} {\bibinfo {title} {Challenges and opportunities in quantum machine learning},\ }\href {https://doi.org/10.1038/s43588-022-00311-3} {\bibfield  {journal} {\bibinfo  {journal} {Nature Computational Science}\ }\textbf {\bibinfo {volume} {2}},\ \bibinfo {pages} {567} (\bibinfo {year} {2022})}\BibitemShut {NoStop}%
\bibitem [{\citenamefont {Meyer}\ \emph {et~al.}(2023)\citenamefont {Meyer}, \citenamefont {Mularski}, \citenamefont {Gil-Fuster}, \citenamefont {Mele}, \citenamefont {Arzani}, \citenamefont {Wilms},\ and\ \citenamefont {Eisert}}]{PRXQuantum.4.010328}%
  \BibitemOpen
  \bibfield  {author} {\bibinfo {author} {\bibfnamefont {J.~J.}\ \bibnamefont {Meyer}}, \bibinfo {author} {\bibfnamefont {M.}~\bibnamefont {Mularski}}, \bibinfo {author} {\bibfnamefont {E.}~\bibnamefont {Gil-Fuster}}, \bibinfo {author} {\bibfnamefont {A.~A.}\ \bibnamefont {Mele}}, \bibinfo {author} {\bibfnamefont {F.}~\bibnamefont {Arzani}}, \bibinfo {author} {\bibfnamefont {A.}~\bibnamefont {Wilms}},\ and\ \bibinfo {author} {\bibfnamefont {J.}~\bibnamefont {Eisert}},\ }\bibfield  {title} {\bibinfo {title} {Exploiting symmetry in variational quantum machine learning},\ }\href {https://doi.org/10.1103/PRXQuantum.4.010328} {\bibfield  {journal} {\bibinfo  {journal} {PRX Quantum}\ }\textbf {\bibinfo {volume} {4}},\ \bibinfo {pages} {010328} (\bibinfo {year} {2023})}\BibitemShut {NoStop}%
\bibitem [{\citenamefont {Schatzki}\ \emph {et~al.}(2024)\citenamefont {Schatzki}, \citenamefont {Larocca}, \citenamefont {Nguyen}, \citenamefont {Sauvage},\ and\ \citenamefont {Cerezo}}]{Schatzki2024}%
  \BibitemOpen
  \bibfield  {author} {\bibinfo {author} {\bibfnamefont {L.}~\bibnamefont {Schatzki}}, \bibinfo {author} {\bibfnamefont {M.}~\bibnamefont {Larocca}}, \bibinfo {author} {\bibfnamefont {Q.~T.}\ \bibnamefont {Nguyen}}, \bibinfo {author} {\bibfnamefont {F.}~\bibnamefont {Sauvage}},\ and\ \bibinfo {author} {\bibfnamefont {M.}~\bibnamefont {Cerezo}},\ }\bibfield  {title} {\bibinfo {title} {Theoretical guarantees for permutation-equivariant quantum neural networks},\ }\href {https://doi.org/10.1038/s41534-024-00804-1} {\bibfield  {journal} {\bibinfo  {journal} {npj Quantum Information}\ }\textbf {\bibinfo {volume} {10}},\ \bibinfo {pages} {12} (\bibinfo {year} {2024})}\BibitemShut {NoStop}%
\bibitem [{\citenamefont {Le}\ \emph {et~al.}(2023)\citenamefont {Le}, \citenamefont {Kiss}, \citenamefont {Schuhmacher}, \citenamefont {Tavernelli},\ and\ \citenamefont {Tacchino}}]{le2023symmetryinvariantquantummachinelearning}%
  \BibitemOpen
  \bibfield  {author} {\bibinfo {author} {\bibfnamefont {I.~N.~M.}\ \bibnamefont {Le}}, \bibinfo {author} {\bibfnamefont {O.}~\bibnamefont {Kiss}}, \bibinfo {author} {\bibfnamefont {J.}~\bibnamefont {Schuhmacher}}, \bibinfo {author} {\bibfnamefont {I.}~\bibnamefont {Tavernelli}},\ and\ \bibinfo {author} {\bibfnamefont {F.}~\bibnamefont {Tacchino}},\ }\href {https://arxiv.org/abs/2311.11362} {\bibinfo {title} {Symmetry-invariant quantum machine learning force fields}} (\bibinfo {year} {2023}),\ \Eprint {https://arxiv.org/abs/2311.11362} {arXiv:2311.11362 [quant-ph]} \BibitemShut {NoStop}%
\bibitem [{\citenamefont {Kawai}\ and\ \citenamefont {Nakagawa}(2020)}]{ClassicalShadowExcitedEnergies}%
  \BibitemOpen
  \bibfield  {author} {\bibinfo {author} {\bibfnamefont {H.}~\bibnamefont {Kawai}}\ and\ \bibinfo {author} {\bibfnamefont {Y.~O.}\ \bibnamefont {Nakagawa}},\ }\bibfield  {title} {\bibinfo {title} {Predicting excited states from ground state wavefunction by supervised quantum machine learning},\ }\href {https://doi.org/10.1088/2632-2153/aba183} {\bibfield  {journal} {\bibinfo  {journal} {Machine Learning: Science and Technology}\ }\textbf {\bibinfo {volume} {1}},\ \bibinfo {pages} {045027} (\bibinfo {year} {2020})}\BibitemShut {NoStop}%
\bibitem [{\citenamefont {Yao}\ \emph {et~al.}(2024)\citenamefont {Yao}, \citenamefont {Ji}, \citenamefont {Li}, \citenamefont {Zhang}, \citenamefont {Chen}, \citenamefont {Ju}, \citenamefont {Liu},\ and\ \citenamefont {Wang}}]{DNNForExcitedStates}%
  \BibitemOpen
  \bibfield  {author} {\bibinfo {author} {\bibfnamefont {Q.}~\bibnamefont {Yao}}, \bibinfo {author} {\bibfnamefont {Q.}~\bibnamefont {Ji}}, \bibinfo {author} {\bibfnamefont {X.}~\bibnamefont {Li}}, \bibinfo {author} {\bibfnamefont {Y.}~\bibnamefont {Zhang}}, \bibinfo {author} {\bibfnamefont {X.}~\bibnamefont {Chen}}, \bibinfo {author} {\bibfnamefont {M.-G.}\ \bibnamefont {Ju}}, \bibinfo {author} {\bibfnamefont {J.}~\bibnamefont {Liu}},\ and\ \bibinfo {author} {\bibfnamefont {J.}~\bibnamefont {Wang}},\ }\bibfield  {title} {\bibinfo {title} {Machine learning accelerates precise excited-state potential energy surface calculations on a quantum computer},\ }\href {https://doi.org/10.1021/acs.jpclett.4c01445} {\bibfield  {journal} {\bibinfo  {journal} {The Journal of Physical Chemistry Letters}\ }\textbf {\bibinfo {volume} {15}},\ \bibinfo {pages} {7061} (\bibinfo {year} {2024})}\BibitemShut {NoStop}%
\bibitem [{\citenamefont {Jordan}\ and\ \citenamefont {Wigner}(1928)}]{JW_transformation}%
  \BibitemOpen
  \bibfield  {author} {\bibinfo {author} {\bibfnamefont {P.}~\bibnamefont {Jordan}}\ and\ \bibinfo {author} {\bibfnamefont {E.}~\bibnamefont {Wigner}},\ }\bibfield  {title} {\bibinfo {title} {ber das paulische quivalenzverbot},\ }\href {https://doi.org/10.1007/BF01331938} {\bibfield  {journal} {\bibinfo  {journal} {Zeitschrift fr Physik}\ }\textbf {\bibinfo {volume} {47}},\ \bibinfo {pages} {631} (\bibinfo {year} {1928})}\BibitemShut {NoStop}%
\bibitem [{\citenamefont {Vatan}\ and\ \citenamefont {Williams}(2004)}]{optimal_two_qubit_gate}%
  \BibitemOpen
  \bibfield  {author} {\bibinfo {author} {\bibfnamefont {F.}~\bibnamefont {Vatan}}\ and\ \bibinfo {author} {\bibfnamefont {C.}~\bibnamefont {Williams}},\ }\bibfield  {title} {\bibinfo {title} {Optimal quantum circuits for general two-qubit gates},\ }\href {https://doi.org/10.1103/PhysRevA.69.032315} {\bibfield  {journal} {\bibinfo  {journal} {Phys. Rev. A}\ }\textbf {\bibinfo {volume} {69}},\ \bibinfo {pages} {032315} (\bibinfo {year} {2004})}\BibitemShut {NoStop}%
\bibitem [{\citenamefont {Cong}\ \emph {et~al.}(2019)\citenamefont {Cong}, \citenamefont {Choi},\ and\ \citenamefont {Lukin}}]{QCNN}%
  \BibitemOpen
  \bibfield  {author} {\bibinfo {author} {\bibfnamefont {I.}~\bibnamefont {Cong}}, \bibinfo {author} {\bibfnamefont {S.}~\bibnamefont {Choi}},\ and\ \bibinfo {author} {\bibfnamefont {M.~D.}\ \bibnamefont {Lukin}},\ }\bibfield  {title} {\bibinfo {title} {Quantum convolutional neural networks},\ }\href {https://doi.org/10.1038/s41567-019-0648-8} {\bibfield  {journal} {\bibinfo  {journal} {Nature Physics}\ }\textbf {\bibinfo {volume} {15}},\ \bibinfo {pages} {1273} (\bibinfo {year} {2019})}\BibitemShut {NoStop}%
\bibitem [{\citenamefont {Burton}(2024)}]{fixed_adapt_vqe_disso_ansatz}%
  \BibitemOpen
  \bibfield  {author} {\bibinfo {author} {\bibfnamefont {H.~G.~A.}\ \bibnamefont {Burton}},\ }\href {https://arxiv.org/abs/2312.09761} {\bibinfo {title} {Accurate and gate-efficient quantum ans\"atze for electronic states without adaptive optimisation}} (\bibinfo {year} {2024}),\ \Eprint {https://arxiv.org/abs/2312.09761} {arXiv:2312.09761 [physics.chem-ph]} \BibitemShut {NoStop}%
\bibitem [{\citenamefont {Bermejo}\ \emph {et~al.}(2024)\citenamefont {Bermejo}, \citenamefont {Braccia}, \citenamefont {Rudolph}, \citenamefont {Holmes}, \citenamefont {Cincio},\ and\ \citenamefont {Cerezo}}]{QCNN_simulatable}%
  \BibitemOpen
  \bibfield  {author} {\bibinfo {author} {\bibfnamefont {P.}~\bibnamefont {Bermejo}}, \bibinfo {author} {\bibfnamefont {P.}~\bibnamefont {Braccia}}, \bibinfo {author} {\bibfnamefont {M.~S.}\ \bibnamefont {Rudolph}}, \bibinfo {author} {\bibfnamefont {Z.}~\bibnamefont {Holmes}}, \bibinfo {author} {\bibfnamefont {L.}~\bibnamefont {Cincio}},\ and\ \bibinfo {author} {\bibfnamefont {M.}~\bibnamefont {Cerezo}},\ }\href@noop {} {\bibinfo {title} {\href{https://arxiv.org/abs/2408.12739}{Quantum Convolutional Neural Networks are (Effectively) Classically Simulable}}} (\bibinfo {year} {2024})\BibitemShut {NoStop}%
\bibitem [{\citenamefont {Kingma}\ and\ \citenamefont {Ba}(2017)}]{adam}%
  \BibitemOpen
  \bibfield  {author} {\bibinfo {author} {\bibfnamefont {D.~P.}\ \bibnamefont {Kingma}}\ and\ \bibinfo {author} {\bibfnamefont {J.}~\bibnamefont {Ba}},\ }\href@noop {} {\bibinfo {title} {\href{https://arxiv.org/abs/1412.6980}{Adam: A Method for Stochastic Optimization}}} (\bibinfo {year} {2017})\BibitemShut {NoStop}%
\bibitem [{\citenamefont {Sun}\ \emph {et~al.}(2017{\natexlab{b}})\citenamefont {Sun}, \citenamefont {Berkelbach}, \citenamefont {Blunt}, \citenamefont {Booth}, \citenamefont {Guo}, \citenamefont {Li}, \citenamefont {Liu}, \citenamefont {McClain}, \citenamefont {Sayfutyarova}, \citenamefont {Sharma}, \citenamefont {Wouters},\ and\ \citenamefont {Chan}}]{pyscf}%
  \BibitemOpen
  \bibfield  {author} {\bibinfo {author} {\bibfnamefont {Q.}~\bibnamefont {Sun}}, \bibinfo {author} {\bibfnamefont {T.~C.}\ \bibnamefont {Berkelbach}}, \bibinfo {author} {\bibfnamefont {N.~S.}\ \bibnamefont {Blunt}}, \bibinfo {author} {\bibfnamefont {G.~H.}\ \bibnamefont {Booth}}, \bibinfo {author} {\bibfnamefont {S.}~\bibnamefont {Guo}}, \bibinfo {author} {\bibfnamefont {Z.}~\bibnamefont {Li}}, \bibinfo {author} {\bibfnamefont {J.}~\bibnamefont {Liu}}, \bibinfo {author} {\bibfnamefont {J.}~\bibnamefont {McClain}}, \bibinfo {author} {\bibfnamefont {E.~R.}\ \bibnamefont {Sayfutyarova}}, \bibinfo {author} {\bibfnamefont {S.}~\bibnamefont {Sharma}}, \bibinfo {author} {\bibfnamefont {S.}~\bibnamefont {Wouters}},\ and\ \bibinfo {author} {\bibfnamefont {G.~K.-L.}\ \bibnamefont {Chan}},\ }\href@noop {} {\bibinfo {title} {\href{https://arxiv.org/abs/1701.08223}{The Python-based Simulations of Chemistry Framework (PySCF)}}} (\bibinfo {year} {2017}{\natexlab{b}})\BibitemShut {NoStop}%
\bibitem [{\citenamefont {Javadi-Abhari}\ \emph {et~al.}(2024)\citenamefont {Javadi-Abhari}, \citenamefont {Treinish}, \citenamefont {Krsulich}, \citenamefont {Wood}, \citenamefont {Lishman}, \citenamefont {Gacon}, \citenamefont {Martiel}, \citenamefont {Nation}, \citenamefont {Bishop}, \citenamefont {Cross}, \citenamefont {Johnson},\ and\ \citenamefont {Gambetta}}]{qiskit}%
  \BibitemOpen
  \bibfield  {author} {\bibinfo {author} {\bibfnamefont {A.}~\bibnamefont {Javadi-Abhari}}, \bibinfo {author} {\bibfnamefont {M.}~\bibnamefont {Treinish}}, \bibinfo {author} {\bibfnamefont {K.}~\bibnamefont {Krsulich}}, \bibinfo {author} {\bibfnamefont {C.~J.}\ \bibnamefont {Wood}}, \bibinfo {author} {\bibfnamefont {J.}~\bibnamefont {Lishman}}, \bibinfo {author} {\bibfnamefont {J.}~\bibnamefont {Gacon}}, \bibinfo {author} {\bibfnamefont {S.}~\bibnamefont {Martiel}}, \bibinfo {author} {\bibfnamefont {P.~D.}\ \bibnamefont {Nation}}, \bibinfo {author} {\bibfnamefont {L.~S.}\ \bibnamefont {Bishop}}, \bibinfo {author} {\bibfnamefont {A.~W.}\ \bibnamefont {Cross}}, \bibinfo {author} {\bibfnamefont {B.~R.}\ \bibnamefont {Johnson}},\ and\ \bibinfo {author} {\bibfnamefont {J.~M.}\ \bibnamefont {Gambetta}},\ }\href@noop {} {\bibinfo {title} {\href{https://arxiv.org/abs/2405.08810}{Quantum computing with Qiskit}}} (\bibinfo {year} {2024})\BibitemShut {NoStop}%
\bibitem [{\citenamefont {Ansel}\ \emph {et~al.}(2024)\citenamefont {Ansel}, \citenamefont {Yang}, \citenamefont {He}, \citenamefont {Gimelshein}, \citenamefont {Jain}, \citenamefont {Voznesensky}, \citenamefont {Bao}, \citenamefont {Bell}, \citenamefont {Berard}, \citenamefont {Burovski} \emph {et~al.}}]{pytorch2024}%
  \BibitemOpen
  \bibfield  {author} {\bibinfo {author} {\bibfnamefont {J.}~\bibnamefont {Ansel}}, \bibinfo {author} {\bibfnamefont {E.}~\bibnamefont {Yang}}, \bibinfo {author} {\bibfnamefont {H.}~\bibnamefont {He}}, \bibinfo {author} {\bibfnamefont {N.}~\bibnamefont {Gimelshein}}, \bibinfo {author} {\bibfnamefont {A.}~\bibnamefont {Jain}}, \bibinfo {author} {\bibfnamefont {M.}~\bibnamefont {Voznesensky}}, \bibinfo {author} {\bibfnamefont {B.}~\bibnamefont {Bao}}, \bibinfo {author} {\bibfnamefont {P.}~\bibnamefont {Bell}}, \bibinfo {author} {\bibfnamefont {D.}~\bibnamefont {Berard}}, \bibinfo {author} {\bibfnamefont {E.}~\bibnamefont {Burovski}}, \emph {et~al.},\ }\href {https://doi.org/10.1145/3620665.3640366} {\bibinfo {title} {Pytorch 2: Faster machine learning through dynamic python bytecode transformation and graph compilation}} (\bibinfo {year} {2024})\BibitemShut {NoStop}%
\bibitem [{\citenamefont {Kreplin}\ \emph {et~al.}(2023)\citenamefont {Kreplin}, \citenamefont {Willmann}, \citenamefont {Schnabel}, \citenamefont {Rapp}, \citenamefont {Hagel\"uken},\ and\ \citenamefont {Roth}}]{squlearn2023}%
  \BibitemOpen
  \bibfield  {author} {\bibinfo {author} {\bibfnamefont {D.~A.}\ \bibnamefont {Kreplin}}, \bibinfo {author} {\bibfnamefont {M.}~\bibnamefont {Willmann}}, \bibinfo {author} {\bibfnamefont {J.}~\bibnamefont {Schnabel}}, \bibinfo {author} {\bibfnamefont {F.}~\bibnamefont {Rapp}}, \bibinfo {author} {\bibfnamefont {M.}~\bibnamefont {Hagel\"uken}},\ and\ \bibinfo {author} {\bibfnamefont {M.}~\bibnamefont {Roth}},\ }\href {https://doi.org/10.48550/arXiv.2311.08990} {\bibinfo {title} {squlearn - a python library for quantum machine learning}} (\bibinfo {year} {2023})\BibitemShut {NoStop}%
\bibitem [{\citenamefont {Bergholm}\ \emph {et~al.}(2022)\citenamefont {Bergholm}, \citenamefont {Izaac}, \citenamefont {Schuld}, \citenamefont {Gogolin}, \citenamefont {Ahmed}, \citenamefont {Ajith}, \citenamefont {Alam}, \citenamefont {Alonso-Linaje}, \citenamefont {AkashNarayanan}, \citenamefont {Asadi} \emph {et~al.}}]{pennylane}%
  \BibitemOpen
  \bibfield  {author} {\bibinfo {author} {\bibfnamefont {V.}~\bibnamefont {Bergholm}}, \bibinfo {author} {\bibfnamefont {J.}~\bibnamefont {Izaac}}, \bibinfo {author} {\bibfnamefont {M.}~\bibnamefont {Schuld}}, \bibinfo {author} {\bibfnamefont {C.}~\bibnamefont {Gogolin}}, \bibinfo {author} {\bibfnamefont {S.}~\bibnamefont {Ahmed}}, \bibinfo {author} {\bibfnamefont {V.}~\bibnamefont {Ajith}}, \bibinfo {author} {\bibfnamefont {M.~S.}\ \bibnamefont {Alam}}, \bibinfo {author} {\bibfnamefont {G.}~\bibnamefont {Alonso-Linaje}}, \bibinfo {author} {\bibfnamefont {B.}~\bibnamefont {AkashNarayanan}}, \bibinfo {author} {\bibfnamefont {A.}~\bibnamefont {Asadi}}, \emph {et~al.},\ }\href@noop {} {\bibinfo {title} {\href{https://arxiv.org/abs/1811.04968}{PennyLane: Automatic differentiation of hybrid quantum-classical computations}}} (\bibinfo {year} {2022})\BibitemShut {NoStop}%
\bibitem [{\citenamefont {Pedregosa}\ \emph {et~al.}(2011)\citenamefont {Pedregosa}, \citenamefont {Varoquaux}, \citenamefont {Gramfort}, \citenamefont {Michel}, \citenamefont {Thirion}, \citenamefont {Grisel}, \citenamefont {Blondel}, \citenamefont {Prettenhofer}, \citenamefont {Weiss}, \citenamefont {Dubourg}, \citenamefont {Vanderplas}, \citenamefont {Passos}, \citenamefont {Cournapeau}, \citenamefont {Brucher}, \citenamefont {Perrot},\ and\ \citenamefont {Duchesnay}}]{scikit_learn}%
  \BibitemOpen
  \bibfield  {author} {\bibinfo {author} {\bibfnamefont {F.}~\bibnamefont {Pedregosa}}, \bibinfo {author} {\bibfnamefont {G.}~\bibnamefont {Varoquaux}}, \bibinfo {author} {\bibfnamefont {A.}~\bibnamefont {Gramfort}}, \bibinfo {author} {\bibfnamefont {V.}~\bibnamefont {Michel}}, \bibinfo {author} {\bibfnamefont {B.}~\bibnamefont {Thirion}}, \bibinfo {author} {\bibfnamefont {O.}~\bibnamefont {Grisel}}, \bibinfo {author} {\bibfnamefont {M.}~\bibnamefont {Blondel}}, \bibinfo {author} {\bibfnamefont {P.}~\bibnamefont {Prettenhofer}}, \bibinfo {author} {\bibfnamefont {R.}~\bibnamefont {Weiss}}, \bibinfo {author} {\bibfnamefont {V.}~\bibnamefont {Dubourg}}, \bibinfo {author} {\bibfnamefont {J.}~\bibnamefont {Vanderplas}}, \bibinfo {author} {\bibfnamefont {A.}~\bibnamefont {Passos}}, \bibinfo {author} {\bibfnamefont {D.}~\bibnamefont {Cournapeau}}, \bibinfo {author} {\bibfnamefont {M.}~\bibnamefont {Brucher}}, \bibinfo {author} {\bibfnamefont {M.}~\bibnamefont {Perrot}},\ and\ \bibinfo {author} {\bibfnamefont
  {E.}~\bibnamefont {Duchesnay}},\ }\bibfield  {title} {\bibinfo {title} {Scikit-learn: Machine learning in {P}ython},\ }\href@noop {} {\bibfield  {journal} {\bibinfo  {journal} {Journal of Machine Learning Research}\ }\textbf {\bibinfo {volume} {12}},\ \bibinfo {pages} {2825} (\bibinfo {year} {2011})}\BibitemShut {NoStop}%
\bibitem [{\citenamefont {Bowles}\ \emph {et~al.}(2024)\citenamefont {Bowles}, \citenamefont {Ahmed},\ and\ \citenamefont {Schuld}}]{bowles2024betterclassicalsubtleart}%
  \BibitemOpen
  \bibfield  {author} {\bibinfo {author} {\bibfnamefont {J.}~\bibnamefont {Bowles}}, \bibinfo {author} {\bibfnamefont {S.}~\bibnamefont {Ahmed}},\ and\ \bibinfo {author} {\bibfnamefont {M.}~\bibnamefont {Schuld}},\ }\href@noop {} {\bibinfo {title} {\href{https://arxiv.org/abs/2403.07059}{Better than classical? The subtle art of benchmarking quantum machine learning models}}} (\bibinfo {year} {2024})\BibitemShut {NoStop}%
\bibitem [{\citenamefont {Huang}(2022)}]{classical_shadows}%
  \BibitemOpen
  \bibfield  {author} {\bibinfo {author} {\bibfnamefont {H.-Y.}\ \bibnamefont {Huang}},\ }\bibfield  {title} {\bibinfo {title} {Learning quantum states from their classical shadows},\ }\href {https://doi.org/10.1038/s42254-021-00411-5} {\bibfield  {journal} {\bibinfo  {journal} {Nature Reviews Physics}\ }\textbf {\bibinfo {volume} {4}},\ \bibinfo {pages} {81} (\bibinfo {year} {2022})}\BibitemShut {NoStop}%
\bibitem [{\citenamefont {Kreplin}\ and\ \citenamefont {Roth}(2024)}]{Kreplin2024reductionoffinite}%
  \BibitemOpen
  \bibfield  {author} {\bibinfo {author} {\bibfnamefont {D.~A.}\ \bibnamefont {Kreplin}}\ and\ \bibinfo {author} {\bibfnamefont {M.}~\bibnamefont {Roth}},\ }\bibfield  {title} {\bibinfo {title} {Reduction of finite sampling noise in quantum neural networks},\ }\href {https://doi.org/10.22331/q-2024-06-25-1385} {\bibfield  {journal} {\bibinfo  {journal} {{Quantum}}\ }\textbf {\bibinfo {volume} {8}},\ \bibinfo {pages} {1385} (\bibinfo {year} {2024})}\BibitemShut {NoStop}%
\end{thebibliography}%

\appendix

\section{\label{sec:appendix_QNN_parameter_scaling}QNN Parameter Scaling}

In this section, we show that the parameters in the QNN scale linearly with the number of orbitals $n_{\rm orb}$, i.e, $n_{\mathrm{params}} = 8(n_{\mathrm{orb}}-1)-3$. For this we need to note that the first layer consists of $n_{\mathrm{orb}}$ gates, which we call $N$ here, each of which contain three parameters. The subsequent pooling layer consists of $\frac{n_{\mathrm{orb}}}{2}$ gates, called $P$ here, with two parameters each. The next set of $N$ and $P$ gates act on half the qubits and therefore their number of parameters is half the number of parameters of the first layer and so on. In the last layer the circular application of the $N$ gate is not necessary since only two qubits are left each and therefore there are three parameters less in the last layer of $N$ and $P$ gates, which otherwise would consist of eight parameters.

\begin{align}
n_{\mathrm{params}}+3 &= 3 n_{\mathrm{orb}}+n_{\mathrm{orb}}+3 \frac{n_{\mathrm{orb}}}{2}+\frac{n_{\mathrm{orb}}}{2}+...+8
\\
&=4 n_{\mathrm{orb}}+2 n_{\mathrm{orb}}+...+8
\\
&=\sum_{i=1}^{\log_2(n_{\mathrm{orb}})} 2^{i+2}=4((\sum_{i=0}^{\log_2(n_{\mathrm{orb}})} 2^i)-1)
\\
&=4(\frac{2^{\log_2(n_{\mathrm{orb}})+1}-1}{2-1}-1) = 4(2 n_{\mathrm{orb}}-2) 
\\
&=8(n_{\mathrm{orb}}-1)
\end{align}

We do not need to pool down to two qubits left in the circuit but can measure earlier and therefore this is in general an upper limit on the number of parameters in the circuit.

\begin{figure*}[t]
\includegraphics[width=1.0\textwidth,clip, trim= 0cm 0cm 0cm 0cm]{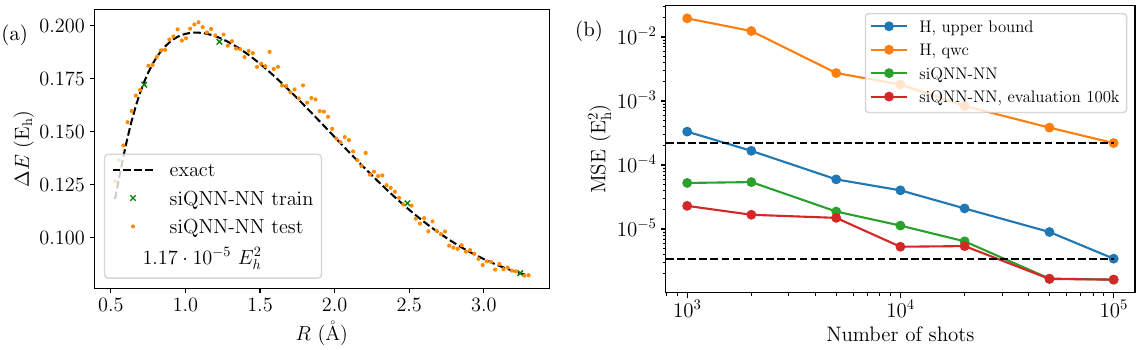}
\caption{\label{fig:results_shot_noise_influence_appendix} The influence of shot noise on the siQNN-NN for a set of training data points and the target function $\Delta E_{T_2}$ for LiH. In (a) the predicted test and train energy differences is shown after training and evaluation with 10,000 shots. In (b) the test MSE depending on the number of shots used for training and evaluation for the siQNN-NN is shown. The \enquote{evaluation 100k} shows the performance when evaluating instead with 100,000 shots. As comparison the MSE on the measurement of the Hamiltonian on the ground state is shown. Here "upper bound" means that each Pauli-string is measured with the given number of shots while \enquote{qwc} means that the given shot budget is shared across qubit wise commuting subsets of the Pauli-strings in the Hamiltonian.}
\end{figure*}

\section{\label{sec:appendix_shot_noise_influence}Influence of Shot Noise}

In current quantum hardware several kinds of noise need to be considered. Due to error correction in the upcoming years the influence of most noise types can be significantly reduced, but shot noise is a challenge intrinsic to quantum computers and will therefore persist beyond the NISQ regime. Therefore in this section we focus on this type of noise and investigate its effects. One of the advantages of the siQNN model is that the number of required shots for a given accuracy is expected to be significantly reduced compared to conventional methods such as VQE extensions. This is because the model directly approximates $\Delta E$ on the quantum state. In order to investigate into this shot reduction we train and test our model for the target function and training data points from Figure~\ref{fig:results_BO-6-31G-LiH_best_example}(a) for different numbers of shots. Figure~\ref{fig:results_shot_noise_influence_appendix}(a) exemplary shows the prediction of the siQNN-NN after training and evaluation with 10.000 shots. We can see that in this case there is no systematic error introduced by the shot noise and that already with $\sim 10^4$ shots the test MSE is lower then the test MSE of the classical comparison models [compare Figure~\ref{fig:results_BO-6-31G-LiH_best_example}(a)].

In Figure~\ref{fig:results_shot_noise_influence_appendix}(b) we show the resulting test MSE's with the given number of shots for the training and test evaluations with the siQNN-NN. The label \enquote{siQNN-NN, evaluation 100k} shows the test MSE when training with the given number of shots and evaluating with 100,000 shots. We show this to differentiate the effect of noise on the training and the evaluation process. This scenario acknowledges that training requires a significantly larger number of model evaluation than inference such that a larger shot budget can be granted at inference.

We compare these results to measuring the exact Hamiltonian (H) on the ground state with different numbers of shots. From this only the ground state energy is derived. To calculate $\Delta E$, the excited state energy would be needed, too. This would only increase the error on $\Delta E$. We, however, neglect this contribution here such that the resulting error can be interpreted as a lower bound for determining $\Delta E$ directly through measurement. The "upper bound" label shows the MSE when measuring each Pauli-string in the Hamiltonian with the given number of shots. In contrast to the observable measured in the siQNN-NN the Hamiltonian consists of many non-commuting Pauli-strings and therefore can not be measured on the same quantum state. In the case shown here the Hamiltonian consists of 276 different Pauli-strings, which can be grouped to 58 qubit wise commuting (qwc) subsets, that can be measured on the same quantum state each. Therefore the total number of shots needs to be shared between these 58 sets, which reduces the number of shots available for each set significantly. The resulting MSE is labeled \enquote{qwc}. More elaborate grouping strategies can be leveraged such as from the classical shadows mechanism to increase the precision but the "upper bound" case is an by far unreachable upper bound to this. Furthermore, many of today's quantum computers use qwc by default.

In this plot we can see that for the given target function and training data points the siQNN-NN derives $\Delta E$ with a significantly lower test MSE then if one would measure with the Hamiltonian on the ground and excited state and subtract to derive $\Delta E$. Even if we neglect the derivation of the excited state energy and the fact that the Hamiltonian consists of many non-commuting Pauli-strings, we need about 100,000 shots to derive $\Delta E$, whereas the siQNN-NN only needs about 30,000 shots in total to achieve a similar accuracy. In a more realistic scenario where the fact that we can only measure commuting sets on the same quantum state is considered we would need about 6 million (a factor of 58) shots for the same accuracy. Therefore our model reduces the number of shots necessary to derive $\Delta E$ for a given accuracy by more then two orders of magnitude relative to traditional methods for the given target function and training data set. Further we can see (\enquote{evaluation 100k}) that due to the shot noise in the training process a varying offset is introduced because of the few training data points considered so that the advantage of sampling the test data with a larger number of shots is not significant.

\begin{figure*}[tb]
\includegraphics[width=1.0\textwidth,clip, trim= 0cm 0cm 0cm 0cm]{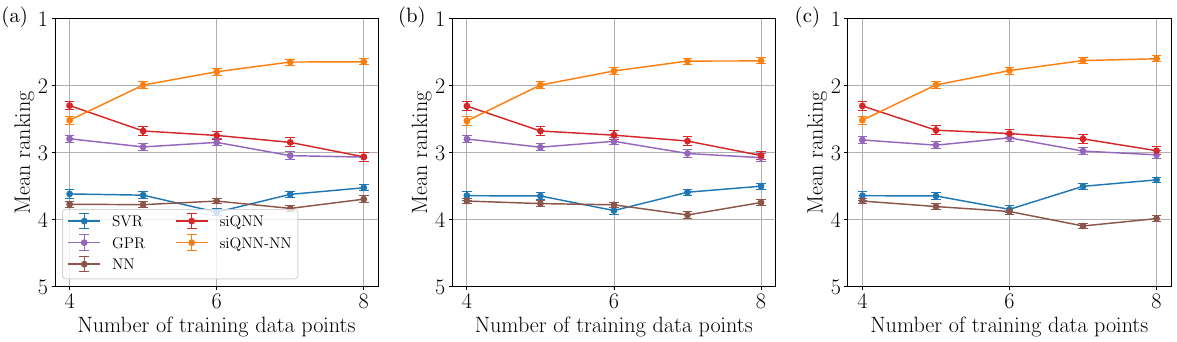}
\caption{\label{fig:results_combined_rankings_plot_appendix} Average rank of the models, broken down by training dataset size for LiH, with all models equal between (a), (b), and (c) except the NN. The number of parameters in the NN corresponds in (a) to 70, in (b) to 150, and in (c) to 1000. Each point represents the relative rank of the respective model compared to all other models averaged over all considered targets and training samples for a molecule. The error bars indicate the standard errors of the mean values of those mean rankings.}
\end{figure*}

\section{\label{sec:larger_NN_comparison}Larger NN comparison}
The small size of the considered NN serves two purposes. Firstly, we want to derive the effect of the NN in the siQNN-NN and therefore need a model with similar size. Secondly, in this low data regime, we consider, one needs to be especially aware of overfitting, which can be reduced by restricting the expressivity of the model (in the example of the NN by reducing its size). In Figure~\ref{fig:results_combined_rankings_plot_appendix} we can see that with an increasing number of parameters in the NN the performance of the NN worsens due to overfitting. We can see that for extremely few training data points (four to five) the effect of overfitting is less relevant. This is due to the fact that in this case many functions can fit the training points with low effort, and no overfitting appears in the pursuit to reach arbitrary precision on the training set.

\section{\label{sec:appendix_mse_plots}MSE Score Plots}
Here, we show the MSE plots derived for the given models and training dataset sizes, on the different molecules and target functions, which are shown in Figure~\ref{fig:method_target_functions}. From the data shown in these MSE plots, the ranking plots in Figure~\ref{fig:results_combined_rankings_plot} are derived.

\begin{figure*}[htbp]
\includegraphics[width=0.96\textwidth,clip, trim= 0cm 0cm 0cm 0cm]{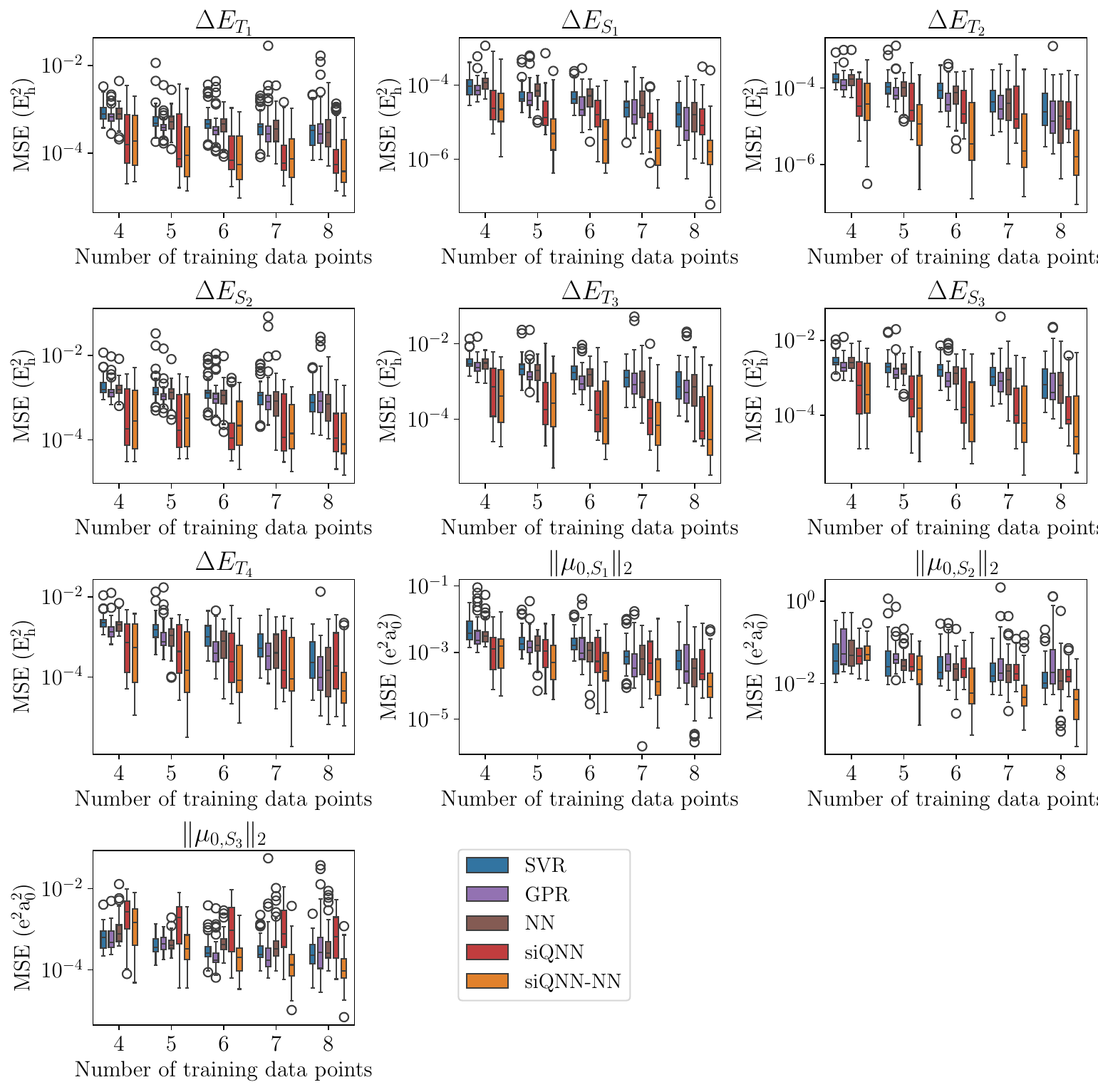}
\caption{\label{fig:appendix_BO-6-31G-LiH} Box plots of the MSE on the test datasets for 50 randomly sampled datasets $\mathcal{D}^L_l$ for the given models and training dataset sizes $L$ for LiH. The heading of each subplot defines the target function, that is fitted.}
\end{figure*}

\begin{figure*}[htbp]
\includegraphics[width=0.96\textwidth,clip, trim= 0cm 0cm 0cm 0cm]{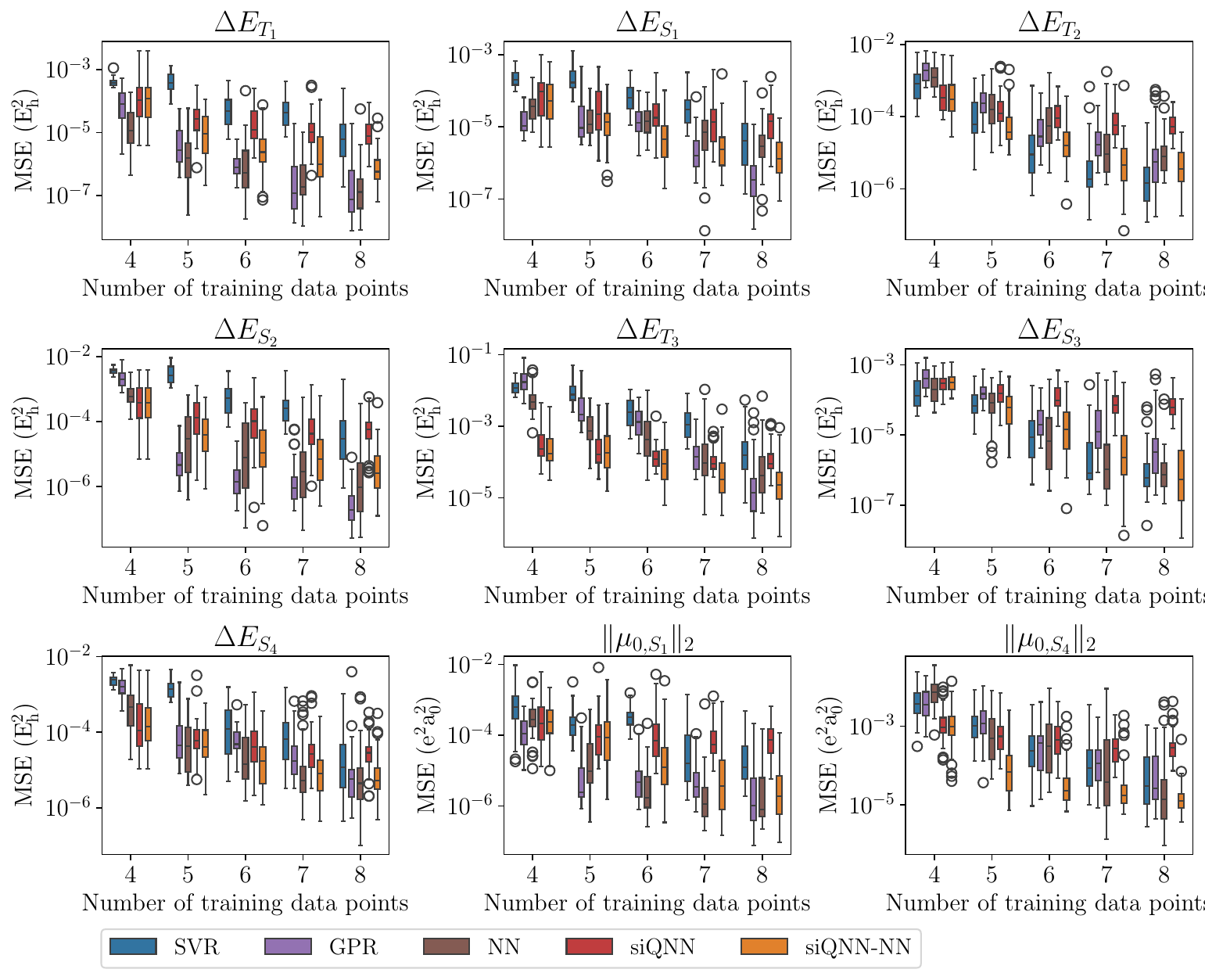}
\caption{\label{fig:appendix_BO-6-31G-H2} Box plots of the MSE on the test datasets for 50 randomly sampled datasets $\mathcal{D}^L_l$ for the given models and training dataset sizes $L$ for $\mathrm{H_2}$. The heading of each subplot defines the target function, that is fitted.}
\end{figure*}

\begin{figure*}[htbp]
\includegraphics[width=0.96\textwidth,clip, trim= 0cm 0cm 0cm 0cm]{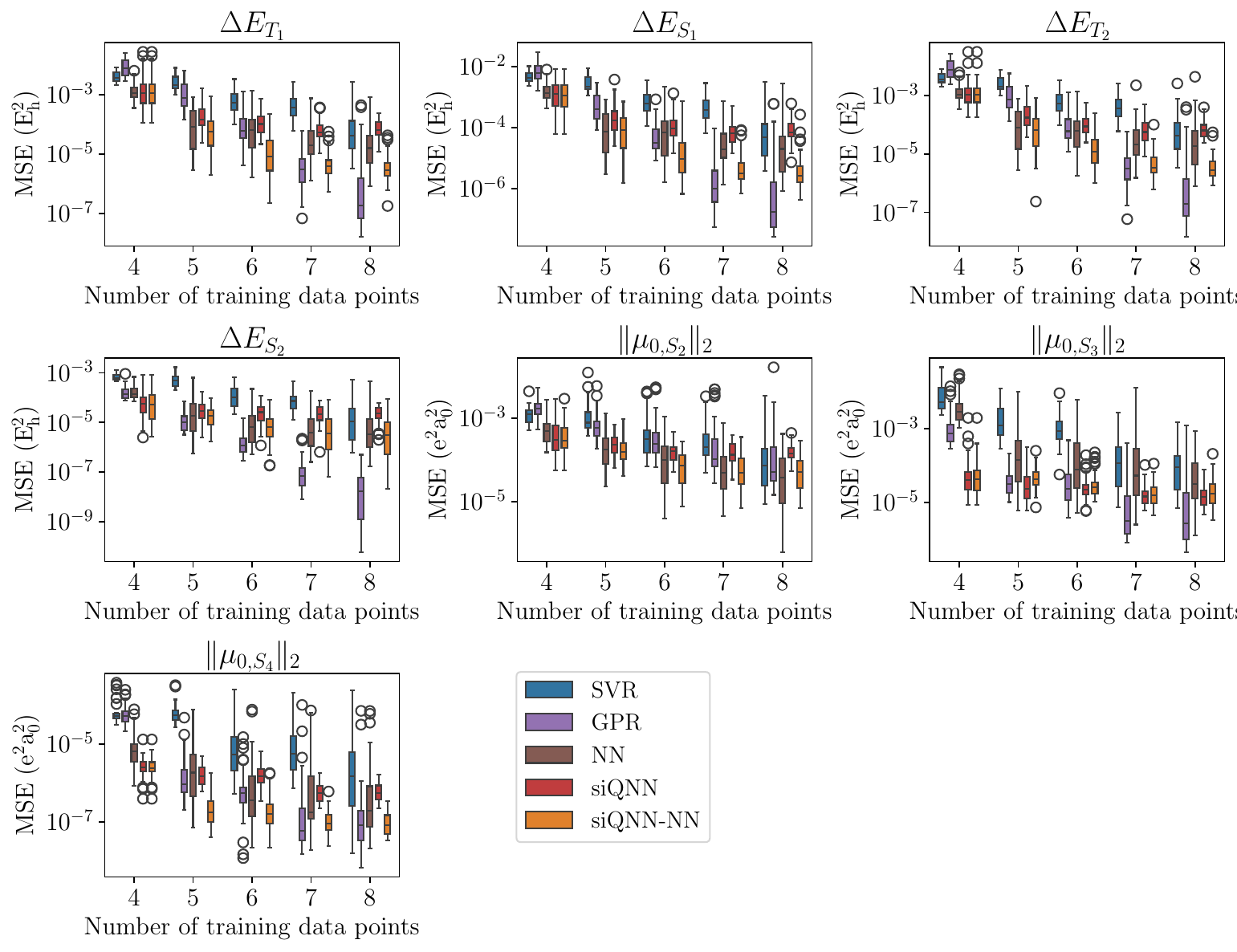}
\caption{\label{fig:appendix_BO-6-31G-H4} Box plots of the MSE on the test datasets for 50 randomly sampled datasets $\mathcal{D}^L_l$ for the given models and training dataset sizes $L$ for $\mathrm{H_4}$. The heading of each subplot defines the target function, that is fitted.}
\end{figure*}

\end{document}